\documentclass[lettersize,journal]{IEEEtran}
\usepackage{amsmath,amsfonts,amssymb}
\usepackage{graphicx}
\usepackage{subfigure}
\usepackage{algorithmicx}
\usepackage{algorithm}
\usepackage{algpseudocode}
\usepackage{textcomp}
\usepackage{makecell}
\usepackage{multirow,multicol}
\usepackage[colorlinks]{hyperref}
\usepackage[table]{xcolor}
\usepackage{threeparttable}

\begin{document}

\title{Multi-Port Selection for FAMA: Massive Connectivity with Fewer RF Chains than Users}

\author{Hanjiang Hong,~\IEEEmembership{Member,~IEEE}, 
        Kai-Kit Wong,~\IEEEmembership{Fellow,~IEEE},\\
        Xusheng Zhu,~\IEEEmembership{Member,~IEEE},
        Hao Xu,~\IEEEmembership{Senior Member,~IEEE},
        Han Xiao,~\IEEEmembership{Student Member,~IEEE},\\
        Farshad Rostami Ghadi,~\IEEEmembership{Member,~IEEE}, and 
        Hyundong Shin, \IEEEmembership{Fellow, IEEE}
\vspace{-7mm}

\thanks{The work of K. K. Wong is supported by the Engineering and Physical Sciences Research Council (EPSRC) under Grant EP/W026813/1.}
\thanks{The work of H. Hong and X. Zhu is supported by the Outstanding Doctoral Graduates Development Scholarship of Shanghai Jiao Tong University.}
\thanks{The work of H. Shin was supported in part by the National Research Foundation of Korea (NRF) grant funded by the Korean government (MSIT) under RS-2025-00556064.}

\thanks{H. Hong, K. K. Wong, X. Zhu, and F. Rostami Ghadi are with the Department of Electronic and Electrical Engineering, University College London, London, United Kingdom. K. K. Wong is also affiliated with the Department of Electronic Engineering, Kyung Hee University, Yongin-si, Gyeonggi-do 17104, Korea (e-mail: \{hanjiang.hong, kai-kit.wong, xusheng.zhu, f.rostamighadi\} @ucl.ac.uk).}
\thanks{H. Xu is with the National Mobile Communications Research Laboratory, Southeast University, Nanjing 210096, China  (e-mail: hao.xu@seu.edu.cn).}
\thanks{H. Xiao is School of Information and Communications Engineering, Xi'an Jiao Tong University, China (e-mail: hanxiaonuli@stu.xjtu.edu.cn).}
\thanks{H. Shin is with the Department of Electronics and Information Convergence Engineering, Kyung Hee University, Yongin-si, Gyeonggi-do 17104, Republic of Korea (e-mail: hshin@khu.ac.kr).}

\thanks{Corresponding author: Kai-Kit Wong.}
}
\maketitle

\begin{abstract}
Fluid antenna multiple access (FAMA) is an emerging technology in massive access designed to meet the demands of future wireless communication networks by naturally mitigating multiuser interference through the utilization of the fluid antenna system (FAS) at RF-chain-limited mobile device. The transition from single-active-port to multi-active-port on a shared RF chain for slow FAMA can greatly enhance its multiplexing capability but is not well understood. Motivated by this, this paper proposes and studies three port selection methods: the optimal exhaustive-search port selection (EPS) as a performance upper bound, and two suboptimal, low-complexity algorithms, namely incremental port selection (IPS) and decremental port selection (DPS). Then the performance of multi-active-port slow FAMA is analyzed, and the complexity of the proposed methods is compared. Simulation results indicate that the proposed methods outperform current state-of-the-art multi-port FAMA techniques. In particular, IPS achieves near-optimal performance while maintaining manageable computational complexity. This research provides a more general framework for port selection in FAMA systems.
\end{abstract}


\vspace{-2mm}
\section{Introduction}
\IEEEPARstart{T}{he relentless} pursuit of even more wireless connectivity driven by the development of sixth-generation (6G) networks and beyond, necessitates revolutionary advancements in physical-layer design \cite{tariq-2020,andrews20246gtakes,ngo2024ultradense}. The explosive growth of the Internet of Things (IoT), massive machine-type communications (mMTC), and artificial intelligence (AI) applications is anticipated to demand extremely massive connectivity at unprecedented data rates with ultra-low latency \cite{nguyen20226ginternet,silva2025distributed}. 

To meet these stringent requirements, researchers are seeking spectrum-sharing techniques that can accommodate a large number of users with sporadic traffic on the same channel-use while managing severe inter-user interference (IUI) and strict latency requirements. In this regard, multiple access schemes focus on efficiently putting users across different dimensions, such as time, frequency, power, space, and code \cite{clercks2024multiple}. In recent years, multiuser multiple-input multiple-output (MU-MIMO) has been the technology to support massive connectivity at the high cost of complexity and power consumption\footnote{When MU-MIMO precoding is used, it will greatly increase the peak-to-average power ratio (PAPR) at each antenna, thereby burdening the power amplifiers and greatly increasing the required power consumption.} \cite{wang2024extremely,pereira2022anOverview}. The latter has become the major hurdle for scaling up precoding to meet the next-generation demands. There are also the recent interest in non-orthogonal multiple access (NOMA) \cite{ahmed2024unveil}, and rate-splitting multiple access (RSMA) \cite{mao2022rsma}. Nonetheless, their aggressive handling of IUI using interference cancellation at the receiver side causes concern and puts doubt on the actual capacity gain when error propagation occurs. For this reason, the search for a massive connectivity scheme continues.

Recently, there was an attempt to avoid precoding and the use of interference cancellation for massive multiple access, motivated by the infrastructure-less massive IoT scenarios. In \cite{wong2022FAMA}, the concept of fluid antenna multiple access (FAMA) was first introduced, which proposes to mitigate interference entirely by repositioning the antenna to where the interference suffers from a deep fade occurred naturally due to multipath fading, without precoding nor interference cancellation. This mechanism relies on the antenna's ability to alter its position on demand, enabled by the fluid antenna system (FAS) concept, envisaged in \cite{wong2020FAS,wong2021FAS}. FAS represents the paradigm of treating antenna as a reconfigurable physical-layer resource to broaden system design for diversity and capacity benefits \cite{New2024aTutorial,Lu-2025,hong2025contemporary,New-2026jsac}. In practice, fluid antennas can come in many forms with different performance-complexity trade-offs, such as liquid-based antennas \cite{shen2024design,Shamim-2025}, metamaterials \cite{Liu-2025arxiv,Zhang-jsac2026}, pixel-based antennas \cite{zhang2024pixel,liu-2025iot}, and others \cite[Sec.~VI]{New2024aTutorial}.

Since its introduction in 2020, many studies have investigated the benefits of FAS over fixed-position antenna systems, see e.g., \cite{Khammassi2023,espinosa2024anew,New2023fluid,new2023information,zhu2025fluid}. Channel state information (CSI) is crucial in determining the optimal position for FAS optimization, which has motivated the research in \cite{xu2023channel,new2025channel,zhang2025successive}, addressing the channel estimation problem in FAS. More recently, positive results were reported regarding its integration into the fifth-generation (5G) New Radio (NR) systems \cite{hong2025fasofdm}.

Turning our attention back to multiple access again, FAMA leverages FAS's unique flexibility to utilize the spatial domain for multiple access. Early FAMA schemes were mainly single-port systems, meaning only one position out of many possible positions is activated to receive the signal \cite{wong2023sFAMA,Xu2024revisiting,wong2022fast,Waqar2023deep}. There are also different types of FAMA schemes:
\begin{itemize}
\item {\bf Fast FAMA} \cite{wong2022FAMA,wong2022fast}---This type requires to switch the port on a symbol-by-symbol basis (or even faster), looking for a per-symbol opportunity where the instantaneous interference signal cancels itself, but this requires rapid channel sensing and identification of such opportunity. 
\item {\bf Slow FAMA} \cite{wong2023sFAMA,Xu2024revisiting,Waqar2023deep}---This type switches the port only once during each channel coherence time, providing a practical and energy-efficient solution.
\item {\bf Coded FAMA} \cite{hong20245gcoded,hong2025coded,waqar2025turbocharging,hong2025Downlink}---This type combines channel coding with the above FAMA techniques.
\end{itemize}
Very briefly, fast FAMA typically has much stronger multiple access capability than slow FAMA but is less practical.

To keep the complexity low as slow FAMA while improving the multiple access capability, the compact ultra-massive array (CUMA) architecture was proposed in \cite{Wong2024cuma}, which activates a large number of ports and aggregate the received signals in the analog domain to produce the output signal for communication. Typically, CUMA considers configurations with only one or two radio frequency (RF) chains at the receiver. It was revealed that CUMA, the multi-port\footnote{In this paper, `multi-port' means activating multiple ports at the same time.} version of slow FAMA, can greatly enhance the IUI rejection capability at each user. Recent work \cite{rao2025geometric} further dived into the optimization of port selection in the CUMA architecture.

While CUMA capitalizes on the analog-domain port activation and combining methodology, especially appealing when the number of RF chains is highly restricted, the 6G edition will see more RF chains at the receiver, prompting the need to consider digital-domain port activation in FAMA systems. In this case, the number of activated ports equals the number of RF chains. In the digital domain, signals from these activated ports are combined by using techniques such as interference rejection combining (IRC) \cite{hong2025Downlink}. IRC optimally weights the received signals to maximize the output signal-to-interference-plus-noise ratio (SINR), thereby providing superior interference mitigation in comparison to single-port selection. 

Preliminary research on digital-domain port activation, however, still uses the single-port selection (SS) method based on the SINR of individual ports \cite{hong20245gcoded,hong2025Downlink}. This neglects the joint correlation among ports, resulting in performance degradation. In \cite{coma2024slow}, a generalized eigenvector (GE)-based scheme was proposed for joint port selection, but its computational complexity renders it prohibitive for FAS with many available ports.


This paper aims to investigate efficient multi-port selection for digital combining in slow FAMA systems. In particular, we propose three distinct port selection methods: exhaustive-search port selection (EPS), incremental port selection (IPS), and decremental port selection (DPS). Through extensive simulations, we demonstrate that the proposed methods achieve a superior trade-off, offering performance improvements over the SS and CUMA techniques. This research serves to provide a framework for efficient multi-port selection in FAMA.

Our main contributions are summarized as follows:
\begin{itemize}
\item First, we propose three port selection methods for multi-port slow FAMA, namely EPS, IPS, and DPS, aimed at maximizing the SINR for the IRC receiver. Specifically, EPS determines the optimal port combination through an exhaustive evaluation of all possibilities, while IPS and DPS construct the port set in a suboptimal incremental or decremental manner. DPS may be regarded as a variation of the GE-based method in \cite{coma2024slow} with a fixed IRC vector. These methods are initially developed using the CSI of the interferers, and subsequently adapted to utilize the estimated covariance matrix of interference-plus-noise, thereby enhancing practical applicability.
\item The performance and complexity of the proposed methods are analyzed. We provide an upper and a lower bound for the average symbol error probability (ASEP) of the multi-port slow FAMA system. In terms of complexity, EPS is only feasible for systems with two RF chains. IPS is suitable for systems with a lower number of RF chains while DPS is preferred for systems where the number of RF chains approaches the number of ports.
\item Simulation results are provided to validate the effectiveness of the proposed methods. At the low frequency band, the proposed techniques surpass both SS and CUMA in performance under finite-scattering channel conditions. But at the high frequency band, the proposed methods are slightly inferior to CUMA with a limited number of RF chains. However, their efficacy converges as the number of RF chains increases. Among the three methodologies, IPS demonstrates near-optimal performance while maintaining manageable complexity. Further simulations over wideband clustered delay line (CDL) channels reveal that in such environments, the proposed methods attain the highest multiplexing gain performance, surpassing even CUMA at the high frequency band.
\end{itemize}

The remainder of this paper is structured as follows: Section \ref{sec:SysMod} introduces the system model for multi-port slow FAMA systems. In Section \ref{sec:PS}, we derive the IRC SINR and present the proposed EPS, IPS, and DPS algorithms. Then Section \ref{sec:Perf} presents numerical results concerning both uncoded and coded performance. Finally, Section \ref{sec:conclusion} concludes the paper.

{\em Notations:} Scalars are represented by lowercase letters while vectors and matrices are denoted by lowercase and uppercase boldface letters, respectively. Transpose and hermitian operations are denoted by superscript $T$ and $\dag$, respectively. For a complex scalar $x$, $\lvert x \rvert$ represents its modulus.

\section{System Model}\label{sec:SysMod}
We consider an interference channel\footnote{This model is valid for uncoordinated multi-cell networks or infrastructure-less device-to-device IoT communication scenarios.} where the base station (BS) communicates to $U$ user terminals (UTs). The BS is equipped with $U$ fixed-position distributed antennas, each responsible for transmitting an information signal to a designated UT. Each UT is equipped with a two-dimensional FAS (2D-FAS) comprising $N=(N_1 \times N_2)$ ports, capable of selecting $N^*$ active ports within a physical size of $W_1\lambda \times W_2 \lambda$, where $\lambda$ is the carrier wavelength. The number of active ports in the system, $N^*$, is dictated by the number of RF chains, $N_{\rm RF}$. Basically, we have $N^* = N_{\rm RF}$. Among the 2D-FAS physical space, $N_i$ ports are uniformly distributed along a linear dimension of length $W_i \lambda$ for $i \in \{1,2\}$. For simplicity, we map the antenna port $(k_1, k_2) \to k: k = (k_1-1)\times N_2 +k_2$, where $k_1 \in \{1, \dots, N_1\}$, $k_2 \in \{1, \dots, N_2\}$, and $k \in \{1, \dots, N\}$. The received signal at the $k$-th port of UT $u$ is given as
\begin{equation}
r_{u,k} [t] = g_{(u,u),k} s_{u}[t] + \sum_{\substack{\tilde{u}=1\\\tilde{u}\neq u}}^{U} g_{(\tilde{u},u),k} s_{\tilde{u}}[t] + \eta_{u,k}[t],
\end{equation}
or in a vector form as
\begin{equation}
\boldsymbol{r}_{u}[t] = \boldsymbol{g}_{(u,u)} s_u[t] + \sum_{\substack{\tilde{u}=1\\\tilde{u}\neq u}}^{U} \boldsymbol{g}_{(\tilde{u},u)} s_{\tilde{u}}[t] + \boldsymbol{\eta}_{u}[t]
\end{equation}
in which $\boldsymbol{r}_{u} = {[r_{u,1},\dots,r_{u,N}]}^T$, $\boldsymbol{\eta}_{u} = {[\eta_{u,1},\dots,\eta_{u,N}]}^T$ with the time index omitted, $\boldsymbol{g}_{(\tilde{u},u)} = {[g_{(\tilde{u},u),1},\dots,g_{(\tilde{u},u),N}]}^T$, $g_{(\tilde{u},u),k}$ denotes the fading channel from the $\tilde{u}$-th BS antenna to UT $u$ at the $k$-th port, $\eta_{u,k}[t]$ represents the zero-mean complex Gaussian noise with variance of $\sigma_\eta^2$, and $s_{u}[t]$ is the transmitted symbol for UT $u$ with $\mathbb{E} [|s_{u}|^2] = \sigma_s^2 = 1$. 

A finite-scattering channel model is adopted so that \cite{buzzi2016clustered}
\begin{multline}
\boldsymbol{g}_{(\tilde{u},u)} =\sqrt{\frac{K\sigma_{(\tilde{u},u)}^2}{K+1}} e^{j\delta_{(\tilde{u},u)}} \boldsymbol{a}\left(\theta_0^{(\tilde{u},u)}, \phi_0^{(\tilde{u},u)}\right)\\ 
+ \sqrt{\frac{\sigma_{(\tilde{u},u)}^2}{N_p(K+1)}} \sum_{l=1}^{N_p} \alpha_l^{(\tilde{u},u)} \boldsymbol{a}\left(\theta_l^{(\tilde{u},u)}, \phi_l^{(\tilde{u},u)}\right),
\end{multline}
where $K$ denotes the Rice factor, $\sigma_{(\tilde{u},u)}^2$ represents the channel power, $\delta_{(\tilde{u},u)}$ signifies the phase of the line-of-sight (LoS) component, $N_p$ indicates the number of scattered components, $\alpha_l^{(\tilde{u},u)}$ is the random complex coefficient of the $l$-th scattered path, and $\boldsymbol{a}(\theta_l^{(\tilde{u},u)}, \phi_l^{(\tilde{u},u)})$ is the steering vector, with $\theta_l^{(\tilde{u},u)}$ and $\phi_l^{(\tilde{u},u)}$ representing the azimuth and elevation angles-of-arrival (AoA), respectively. When $l=0$, the parameters correspond to those for the LoS channel. For simplicity, we set $\sigma_{(\tilde{u},u)} = \sigma, \forall (\tilde{u},u)$ in this paper. Omitting the index $(\tilde{u},u)$, the steering vector $\boldsymbol{a}(\theta_l, \phi_l)$ is defined by
\begin{equation}
\boldsymbol{a}(\theta_l, \phi_l) = {[1,e^{-j\frac{2\pi}{\lambda}d_l(2)},\dots,e^{-j\frac{2\pi}{\lambda}d_l(N)}]}^T,
\end{equation}
where $d_l(k)$ represents the propagation difference of the $l$-th path between the $(1,1)\to 1$-st port and the $(k_1,k_2)\to k$-th port. Accordingly, we have
\begin{equation}
d_l(k) \!=\! \frac{(k_1-1)W_1\lambda}{N_1-1}\sin \theta_l \cos \phi_l + \frac{(k_2-1)W_1\lambda}{N2-1} \cos \theta_l.
\end{equation}

A special case concerns the rich scattering scenario where $K=0$ and $N_p\to \infty$. In this case, the variables, $g_{(\tilde{u},u),k}$, follow an identical complex Gaussian distribution, but the channels ${\{g_{(\tilde{u},u),k}\}}_{\forall k}$ are correlated, characterized by a covariance matrix $ \mathbb{E}\left[ \boldsymbol{g}_{(\tilde{u},u)} \boldsymbol{g}_{(\tilde{u},u)}^ \dag \right] = \boldsymbol{\Sigma}$, defined as \cite{Khammassi2023}
\begin{equation}\label{Eq:corr}
{\left[\boldsymbol{\Sigma}\right]}_{k,\ell}= \sigma^2\! J_0 \!\! \left(\! 2\pi \sqrt{{\left(\frac{k_1-\ell_1}{N_1 - 1} W_1\right)}^2 \!+\! {\left(\frac{k_2-\ell_2}{N_2 - 1} W_2\right)}^2}\right),
\end{equation}
where $J_0(\cdot)$ is the zero-order Bessel function of the first kind.

In conventional slow FAMA systems \cite{wong2023sFAMA}, each UT receiver is equipped with only a single RF chain, and the optimal port that maximizes the received SINR is activated, i.e., $N^* = N_{\rm RF} = 1$. For the $u$-th UT, we have 
\begin{equation}\label{Eq:sFAMA}
    k^* = \arg \max_{k \in \Omega} \frac{\lvert g_{(u,u),k}\rvert ^2}{\sum_{\substack{\tilde{u}=1\\\tilde{u}\neq u}}^{U} \lvert g_{(\tilde{u},u),k}\rvert ^2 + \sigma_\eta^2},
\end{equation}
where $\Omega = \{1,\dots, N\}$ denotes the index set for the ports.

In the cases with multiple RF chains at the UT receiver, CUMA activates many ports $(N^*)$ and aggregates the received signals from these ports in the analog domain into a restricted number of RF chains for digital-domain processing. Consequently, we observe that $N^*\gg N_{\rm RF}$~\cite{Wong2024cuma}. CUMA capitalizes on the combinatorial capability of analog modules. 

Considering that $N^* = N_{\rm RF}>1$, when the selected ports are used in the digital domain, \cite{hong2025Downlink} adopted a multi-port slow FAMA scheme wherein the selection of antenna ports is based on maximizing a single-port SINR. This SS method designates the $n$-th ($n = 1,\dots, N^*$) activated port as
\begin{equation}\label{Eq:SS}
k^*_{n} = \arg \max_{k \in \Omega \backslash \left\{ k^*_1, \dots, k^*_{n-1}\right\}} \frac{\lvert g_{(u,u),k}\rvert ^2}{\sum_{\substack{\tilde{u}=1\\\tilde{u}\neq u}}^{U} \lvert g_{(\tilde{u},u),k}\rvert ^2 + \sigma_\eta^2}.
\end{equation}
Given the index set of the selected ports, denoted as $\kappa = \{k_{u,1}^*,\dots,k^*_{u,N^*}\}$, the received signal vector and the channel vector corresponding to the selected ports are represented as $\hat{\boldsymbol{r}} = {[\boldsymbol{r}]}_\kappa ={[r_{k_1^*}, \dots, r_{k_{N^*}^*}]}^T$ and ${\boldsymbol{h}} = {[\boldsymbol{g}_{(u,u)}]}_\kappa = {[g_{(u,u),k_1^*}, \dots, g_{(u,u),k_{N^*}^*}]}^T$, respectively. The optimal combining vector for IRC, to be applied to $\hat{\boldsymbol{r}}$, is given by 
\begin{equation}
\boldsymbol{w} = {\boldsymbol{h}}^\dag\hat{\boldsymbol{\Phi}}^{-1},
\end{equation}
where $\hat{\boldsymbol{\Phi}}$ represents the covariance matrix of the interference-plus-noise for the selected ports. It encompasses the elements on the $\{\kappa\}$-th rows and $\{\kappa\}$-th columns of $\boldsymbol{\Phi}$, where $\boldsymbol{\Phi}$ is the covariance matrix of all ports, calculated as
\begin{equation}
{\boldsymbol{\Phi}} = \sigma_\eta^2 \boldsymbol{I}_{N} +\sum_{\tilde{u}=1\atop\tilde{u}\neq u}^{U} {\boldsymbol{g}}_{(\tilde{u},u)} {\boldsymbol{g}}_{(\tilde{u},u)}^\dag = \sigma_\eta^2 \boldsymbol{I}_{N} +\boldsymbol{G} \boldsymbol{G}^\dag,
\end{equation}
where $\boldsymbol{G} = {[\boldsymbol{g}_{(1,u)},\dots,\boldsymbol{g}_{(\tilde{u},u)},\dots,\boldsymbol{g}_{(U,u)}]}_{\tilde{u}\neq u} \in \mathbb{C} ^{N\times (U-1)}$ denotes the channel matrix of the interference signals. 

The estimated symbol after combining is thus denoted as $\tilde{s} = \boldsymbol{w}\hat{\boldsymbol{r}}$, and the SINR for the IRC receiver is given by
\begin{equation}\label{Eq:sinr}
\gamma = \boldsymbol{w} \boldsymbol{h} =\boldsymbol{h}^\dag\hat{\boldsymbol{\Phi}}^{-1}\boldsymbol{h}.
\end{equation} 

The SS approach in~\eqref{Eq:SS} does not account for the interference among the other selected ports during the selection process, resulting in a reduction in the IRC receiver's SINR, $\gamma$, and the overall performance. Therefore, the development of port selection methods for multi-port slow FAMA is essential to reliable large-scale connectivity for slow FAMA.

\section{Port Selection Methods}\label{sec:PS}
In this section, we propose three port selection methods for multi-port slow FAMA, namely EPS, IPS, and DPS, with the objective of maximizing SINR as defined in~\eqref{Eq:sinr}. We initially derive a successive SINR expression, followed by the proposal of the three port selection techniques based on this expression, along with an analysis of their performance and computational complexity. The initial formulation of the successive SINR expression is given in Section \ref{subsec:SINRS} and the proposed methods are based on the instantaneous interference CSI, which may have limited practical applicability due to the difficulty in acquiring channel coefficients for interference sources. Consequently, the calculations of SINR are adapted to utilize an estimate of the covariance matrix of interference-plus-noise in Section \ref{subsec:ECM}, which can be more easily estimated.

\subsection{Successively SINR Expression}\label{subsec:SINRS}
The SINR in~\eqref{Eq:sinr} necessitates an inversion of $\hat{\boldsymbol{\Phi}}$, a procedure of high computational complexity. This can be avoided by the application of the matrix inverse lemma. In this subsection, we present an alternative successive fomulation of the SINR with the IRC receiver of the selected ports. This may facilitate a low-complexity calculation of the SINR. 

Let $\boldsymbol{u}_k~(k = 1,\dots,N)$ be the row vectors of $\boldsymbol{G}$, such that $\boldsymbol{G} = {[\boldsymbol{u}_1^T, \dots, \boldsymbol{u}_N^T]}^T$. When selecting $n$ activated ports, with the index set of $\kappa_n = \{k_1,k_2,\dots,k_n\}$, then we have the channel vector of selected ports as $\boldsymbol{h}_n = {[\boldsymbol{g}_{(u,u)}]}_{\kappa_n} = {[g_{(u,u),k_1},\dots, g_{(u,u),k_n}]}^T$, and the interference-plus-noise covariance matrix is $\boldsymbol{\Psi}_n = {[\boldsymbol{\Phi}]}_{\kappa_n, \kappa_n} = \sigma_\eta^2 \boldsymbol{I}_n + \hat{\boldsymbol{G}}_n \hat{\boldsymbol{G}}_n^\dag$, where $\hat{\boldsymbol{G}}_n  = {[\boldsymbol{u}_{k_1}^T,\dots,\boldsymbol{u}_{k_{n}}^T]}^T$. The SINR is then given by
\begin{align}
    \gamma_n \! & \hspace{.5mm}= \boldsymbol{h}_n^\dag {\boldsymbol{\Psi}}_n \boldsymbol{h}_n \label{Eq:sinrD} \\
    & \hspace{.5mm}= \boldsymbol{h}_n^\dag {\left(\sigma_\eta^2 \boldsymbol{I}_n + \hat{\boldsymbol{G}}_n \hat{\boldsymbol{G}}_n^\dag\right)}^{-1} \boldsymbol{h}_n \nonumber\\
    & \overset{(a)}{=} \frac{\|\boldsymbol{h}_n\|^2}{\sigma_\eta^2} \!-\! \frac{1}{\sigma_\eta^4}\boldsymbol{h}_n^\dag \hat{\boldsymbol{G}}_n {\!\left(\boldsymbol{I}_{U\!-\!1} \!+\! \frac{1}{\sigma_\eta^2} \hat{\boldsymbol{G}}_n^\dag \hat{\boldsymbol{G}}_n\right)}^{\!-1} \!\! \hat{\boldsymbol{G}}_n^\dag \boldsymbol{h}_n, \label{Eq:sinrS}
\end{align}
where $(a)$ is obtained from the matrix inverse lemma. Define $\boldsymbol{U}_n = {(\boldsymbol{I}_{U-1}+ \hat{\boldsymbol{G}}_n^\dag \hat{\boldsymbol{G}}_n/\sigma_\eta^2)}^{\!-1}, ~ n = 1,\dots,N^*$. The inversion of $\boldsymbol{U}_n$ is computationally intensive. A viable simplification involves applying the matrix inverse lemma, such that
\begin{align}\label{Eq:Un}
    \boldsymbol{U}_{n} & = {\left(\boldsymbol{I}_{U-1} + \frac{1}{\sigma_\eta^2} \hat{\boldsymbol{G}}_n^\dag \hat{\boldsymbol{G}}_n\right)}^{-1} \nonumber\\
    & = {\left(\boldsymbol{U}_{n-1}^{-1} + \frac{1}{\sigma_\eta^2}\boldsymbol{u}_{k_{n}^*}^\dag \boldsymbol{u}_{k_{n}^*}\right)}^{-1} \nonumber\\
    & = \boldsymbol{U}_{n-1} - \frac{\boldsymbol{U}_{n-1} \boldsymbol{u}_{k_{n}^*}^\dag \boldsymbol{u}_{k_{n}^*} \boldsymbol{U}_{n-1}}{\sigma^2_\eta + \boldsymbol{u}_{k_{n}^*} \boldsymbol{U}_{n-1}\boldsymbol{u}_{k_{n}^*}^\dag}. 
\end{align}
Once $\boldsymbol{U}_{n-1}$ is known, $\boldsymbol{U}_{n}$ can be calculated without performing a matrix inversion. Successively, we can obtain $\boldsymbol{U}_{n}$ in a recursive manner by initiating from $\boldsymbol{U}_0 = \boldsymbol{I}_{U-1}$. 

\subsection{EPS}
In the EPS method, ${N \choose N^*}$ combinations are evaluated, and the one with the highest SINR is selected as the final port set.  For each combination $\omega_i$, the SINR, $\gamma_i$, is calculated either directly or sequentially as detailed in Section~\ref{subsec:SINRS}. 

The direct SINR calculation is given according to~\eqref{Eq:sinrD} by
\begin{equation}\label{Eq:sinrD_EPS}
\gamma_i = \boldsymbol{h}_i^\dag \boldsymbol{\Psi}_i^{-1}\boldsymbol{h}_i,
\end{equation}
where $\boldsymbol{h}_i = {[\boldsymbol{g}_{(u,u)}]}_{\omega_i}$ represents the selected channel vector, comprising of the $\omega_i$-th elements of the channel vector $\boldsymbol{g}_{(u,u)}$, and $\boldsymbol{\Psi}_i = {[\boldsymbol{\Phi}]}_{\omega_i,\omega_i}\in \mathbb{C}^{N^* \times N^*}$ includes the elements located at the $\omega_i$-th rows and $\omega_i$-th columns of $\boldsymbol{\Phi}$.

Based on~\eqref{Eq:sinrS}, the successive SINR calculation becomes
\begin{equation}\label{Eq:sinrS_EPS}
    \gamma_i =  \frac{\|\boldsymbol{h}_i\|^2}{\sigma_\eta^2} -\frac{1}{\sigma_\eta^4}\boldsymbol{h}_i^\dag \dot{\boldsymbol{G}}_i \boldsymbol{U}^{(i)}_{N^*} \dot{\boldsymbol{G}}_i^\dag \boldsymbol{h}_i,
\end{equation}
where $\dot{\boldsymbol{G}}_i\in \mathbb{C} ^{N_{RF}\times (U-1)}$ consists of  $\omega_i$-th row vectors of $\boldsymbol{G}$, $\boldsymbol{U}^{(i)}_{N^*}$ can be iteratively calculated as~\eqref{Eq:Un} with the activation of the port in the index set $\omega_i$ in an incremental manner.
The complete algorithm is summarized in Algorithm~\ref{Alg:EPS}. 
\begin{algorithm} [tbp]
    \caption{Exhaustive-search port selection (EPS)}\label{Alg:EPS}
    \begin{algorithmic}[1]
        \Require{$\boldsymbol{g}_{(u,u)}$, $\boldsymbol{\Phi}$, $N$, $N^*$}
        \Ensure{Index set of selected ports $\kappa$} 
        \State{Initialization: $\Omega = \{1,2,\dots,N\}$, $\gamma = 0$}
        \State{Generate the combinations: $C = {\rm n\_choose\_k}(\Omega, N^*)$}
        \For{$i = 1$ to ${N \choose N^*}$} 
            \State{Choose the $i$-th combination: $\omega_i = C(i,:)$}
            \State{${\boldsymbol{h}}_i = {[\boldsymbol{g}_{(u,u)}]}_{\omega_i}$, ${\boldsymbol{\Psi}}_i = {[\boldsymbol{\Phi}]}_{\omega_i,\omega_i}$}
            \State{Calculate SINR $\gamma_i$ by~\eqref{Eq:sinrD_EPS} or by~\eqref{Eq:sinrS_EPS}} 
            \If{$\gamma_i > \gamma$}
                \State{$\kappa = \omega_i$, $\gamma = \gamma_i$, ${\boldsymbol{h}} = \boldsymbol{h}_i$, $\hat{\boldsymbol{\Phi}} = \boldsymbol{\Psi}_i$}
            \EndIf
        \EndFor
    \end{algorithmic}
\end{algorithm}

We analyze the computational complexity of EPS in terms of the number of complex multiplication operations. The direct SINR calculation requires $C_{\rm D,EPS}^{(i)} = 2N_{\rm RF}^3 + N_{\rm RF}^2 +N_{\rm RF}$ multiplication operations, where the first term, $2N_{\rm RF}^3$, comes from the matrix inverse operation of $\boldsymbol{\Psi}_i^{-1}$, and $N_{\rm RF} = N^*$. In the case of the successive method for calculating SINR, the computation of~\eqref{Eq:Un} requires $(2I^2+I)$ multiplication operations, where $I = U-1$ is the number of interferers. The successive computation involves $N_{\rm RF}$ iterations of~\eqref{Eq:Un} and SINR calculation in~\eqref{Eq:sinrS_EPS} requires $(N_{\rm RF} \!+\! I N_{\rm RF} \!+\! I^2 \!+\! I)$ multiplication operations. Thus, the total number of multiplications for a single SINR calculation is $C_{\rm S,EPS}^{(i)} = {(2N_{\rm RF}+1)(U^2-U)+N_{\rm RF}}$. The EPS algorithm involves ${N \choose N_{\rm RF}}$ iterations, with each iteration having $N_{{\rm EPS}}^{(i)} = 1$. As a result, the overall complexity of the EPS algorithm is estimated to be
\begin{equation}
\mathcal{O}\left({N\choose N_{\rm RF}} \times \min\left\{C_{\rm D,EPS}^{(i)},C_{\rm S,EPS}^{(i)}\right\}\right). 
\end{equation}
When $N_{\rm RF}^2 < U^2-U $, we have $ \min\{C_{\rm D,EPS}^{(i)},C_{\rm S,EPS}^{(i)}\} = C_{\rm D,EPS}^{(i)}$, indicating that the complexity of direct SINR calculation is lower than that of the successive SINR calculation within the EPS method, and vice versa. 

\subsection{IPS}
The IPS method starts with an empty set and sequentially incorporates each activated port into the set. Each addition of a port is determined by a greedy algorithm aimed at maximizing the SINR of the combined signal within the selected set.

In the first iteration, $\omega_{1,k} = \{k\}$, ${h}_{1,k} = g_{(u,u),k}$, ${\Psi}_{1,k} = {[\boldsymbol{\Phi}]}_{k,k}$, and the SINR, $\gamma_{1,k}$, of the $k$-th port for direct SINR calculation is given by
\begin{equation}\label{Eq:sinrD_1_IPS}
    \gamma_{1,k} = {|{h}_{1,k}|^2}/{|{\Psi}_{1,k}|^2}.
\end{equation}
For the successive SINR expression, $\dot{\boldsymbol{G}}_{1,k} = \boldsymbol{u}_k$ is the $k$-th row vector of $\boldsymbol{G}$, and the SINR, $\gamma_{1,k}$, of the selecting $k$-th port can be expressed as 
\begin{equation}\label{Eq:sinrS_1_IPS}
    \gamma_k = \frac{|{h}_{1,k}|^2}{\sigma_\eta^2} - \frac{1}{\sigma_\eta^4} {h}_{1,k}^* \boldsymbol{u}_k \boldsymbol{V}_{1,k} \boldsymbol{u}_k^\dag {h}_{1,k},
\end{equation} 
where
\begin{equation}
\boldsymbol{V}_{\!1,k} \triangleq {\left(\!\boldsymbol{U}_0 \!+\! \frac{1}{\sigma_\eta^2}\boldsymbol{u}_k^\dag \boldsymbol{u}_k\!\!\right)}^{\!-1} = \boldsymbol{I}_{U\!-\!1} \!-\! \boldsymbol{u}_k^\dag {\left(\!\sigma_\eta^2 \!+\! \boldsymbol{u}_k \boldsymbol{u}_k^\dag\!\right)}^{\!-1} \!\!\boldsymbol{u}_k. 
\end{equation}
Thus,~\eqref{Eq:sinrS_1_IPS} can be rewritten as
\begin{align}
    \gamma_k & = \frac{|{h}_{1,k}|^2}{\sigma_\eta^2}\left[1\!-\!\frac{1}{\sigma_\eta^2}\boldsymbol{u}_k \left(\boldsymbol{I}_{U\!-\!1} \!-\! \frac{\boldsymbol{u}_k^\dag \boldsymbol{u}_k}{\sigma_\eta^2 \!+\! \boldsymbol{u}_k \boldsymbol{u}_k^\dag}\right) \boldsymbol{u}_k^\dag\right] \nonumber\\
    & = \frac{|{h}_{1,k}|^2}{\sigma_\eta^2 + \boldsymbol{u}_k\boldsymbol{u}_k^\dag} = \frac{|g_{(u,u),k}|^2}{\sigma_\eta^2 + \sum_{\substack{\tilde{u}=1\\\tilde{u}\neq u}}^{U} \lvert g_{(\tilde{u},u),k}\rvert ^2},
\end{align}
which is identical to~\eqref{Eq:sinrD_1_IPS}. According to~\eqref{Eq:sinrD_1_IPS} or~\eqref{Eq:sinrS_1_IPS}, the first selected port $k_1^*$ in IPS is the one that has the largest single-port SINR, which is the same as in~\eqref{Eq:sFAMA}.

Starting from the second iteration, the $i$-th port can be selected during the $i$-th iteration through comparison of the SINR, $\gamma_{i,k}$, after the IRC combining process. By progressively selecting the $k$-th port as the $i$-th chosen port, the index set is represented as $\omega_{i,k} = \kappa \cup k$ with $k \in \Omega\backslash \{k_1^*\dots, k_{i-1}^*\}$, and the channel vector can be expressed as $\boldsymbol{h}_{i,k} = {[\boldsymbol{g}_{u,u}]}_{\omega_{i,k}}$. For direct SINR calculation, the covariance matrix is derived as $\boldsymbol{\Psi}_{i,k} = {[\boldsymbol{\Phi}]}_{\omega_{i,k}, \omega_{i,k}}\in \mathbb{C}^{i\times i}$, and the combining SINR for the selected ports with index set $\omega_{i,k}$ is given by
\begin{equation}\label{Eq:sinrD_n_IPS}
    \gamma_{i,k} = \boldsymbol{h}_{i,k}^\dag \boldsymbol{\Psi}_{i,k}^{-1} \boldsymbol{h}_{i,k}.
\end{equation}
Using the successive SINR expression, we have $\dot{\boldsymbol{G}}_{i,k} = [\dot{\boldsymbol{G}}_{i-1,k_{i-1}^*}^T, \boldsymbol{u}_k^T]^T$, and $\gamma_{i,k}$ is calculated based on~\eqref{Eq:sinrS} as
\begin{equation}\label{Eq:sinrS_n_IPS}
    \gamma_{i,k} = \frac{\|\boldsymbol{h}_{i,k}\|^2}{\sigma_\eta^2} \!-\! \frac{1}{\sigma_\eta^4}\boldsymbol{h}_{i,k}^\dag \dot{\boldsymbol{G}}_{i,k} \boldsymbol{V}_{i,k} \dot{\boldsymbol{G}}_{i,k}^\dag \boldsymbol{h}_{i,k},
\end{equation}
where we define the matrix $\boldsymbol{V}_{i,k} \triangleq {(\boldsymbol{U}_{i-1}^{-1}+\boldsymbol{u}_k^\dag \boldsymbol{u}_k/\sigma_\eta^2)}^{-1}$, and $\boldsymbol{U}_{i-1} = {(\boldsymbol{U}_{i-2}+\boldsymbol{u}_{k_{i-1}^*}^\dag \boldsymbol{u}_{k_{i-1}^*}/\sigma_\eta^2)}^{-1}$ has been computed during the $(n-1)$-th iteration. 
This iterative process is repeated until $N^* = N_{\rm RF}$ ports are selected. The complete algorithm of IPS is summarized in Algorithm~\ref{Alg:IPS}. 
\begin{algorithm} [tbp]
    \caption{Incremental port selection (IPS)}\label{Alg:IPS}
    \begin{algorithmic}[1]
        \Require{$\boldsymbol{g}_{(u,u)}$, $\boldsymbol{\Phi}$, $N$, $N^*$}
        \Ensure{Index set of selected ports $\kappa$}
        \State{Initialization: $\Omega = \{1,2,\dots,N\}$, $\kappa = \emptyset$}
        \For{$i = 1$ to $N^*$}
            \State{$\omega_{i,k} = \kappa \cup k,~ \forall k \in \Omega$}
            \State{$\boldsymbol{h}_{i,k} = {[\boldsymbol{g}_{(u,u)}]}_{\omega_{i,k}},~\boldsymbol{\Psi}_{i,k} = {[\boldsymbol{\Phi}]}_{\omega_{i,k}, \omega_{i,k}},~ \forall k \in \Omega$} 
            \State{Calculate SINR $\gamma_{i,k}$ by~\eqref{Eq:sinrD_n_IPS} or~\eqref{Eq:sinrS_n_IPS}} 
            \State{$k_i^* = \arg \max_{k \in \Omega} \gamma_{i,k}$} 
            \State{Update $\kappa = \kappa \cup k_i^*$, $\Omega = \Omega \backslash k_i^*$} 
        \EndFor
    \end{algorithmic}
\end{algorithm}

Consider the complexity of the IPS algorithm, with $\boldsymbol{h}_{i,k} \in \mathbb{C}^{i\times 1}$ during the $i$-th iteration. Consequently, the direct SINR calculation requires $C_{{\rm D,IPS}}^{(i)} = 2i^3 + i^2 + i$ multiplication operations, while the number of multiplication for successive SINR calculation is $C_{{\rm S,IPS}}^{(i)} = 3U^2 +(i-4)U +1$. Given that the computation of $\boldsymbol{U}_{n-1}$ was performed during the $(i-1)$-th iteration, the calculation of $\boldsymbol{V}_{i,k}$ requires only $(2U^2-3U+1)$ multiplication operations, thereby reducing the complexity of successive SINR calculations. The total number of iterations in the IPS algorithm is $N_{\rm RF}$. During the $i$-th iteration, the number of SINR calculations is $N_{{\rm IPS}}^{(i)} = N-i+1$. Therefore, the overall computational complexity of the IPS algorithm is expressed as $\mathcal{O}(\sum_{n=1}^{N_{\rm RF}} N_{{\rm IPS}}^{(i)} \times \min\{C_{\rm D,IPS}^{(i)},C_{\rm S,IPS}^{(i)}\})$.

\subsection{DPS}
It is also feasible to select ports by initially choosing all the available ports and subsequently discarding those in a manner that retains ports yielding the highest SINR. This approach is also referred to as GE port selection (GEPort) in~\cite{coma2024slow} wherein the combining vector $\boldsymbol{w}$ is optimized.

The DPS method commences by initializing the port index set as $\Omega = \{1,\dots,N\}$. During the $i$-th iteration, the temporary index set is selected as $\omega_{i,k} = \Omega \backslash k, \forall k\in \Omega\backslash\{k_1\dots,k_{i-1}\}$. With the channel vector $\boldsymbol{h}_{i,k} = {[g_{(u,u),k}]}_{\omega_{i,k}}$, and the covariance matrix $\boldsymbol{\Psi}_{i,k} = {[\boldsymbol{\Phi}]}_{\omega_{i,k},\omega_{i,k}}$, the combining SINR of the ports within set of $\omega_{i,k}$ can be directly calculated as in~\eqref{Eq:sinrD_n_IPS}. For successive SINR computation, let $\ddot{\boldsymbol{G}}_{i,k}$ be the matrix comprising the row vectors of $\boldsymbol{G}$ with indices belonging to the set $\omega_{i,k}$. Then the SINR, $\gamma_{i,k}$, is given by
\begin{equation}\label{Eq:sinrS_DPS}
    \gamma_{i,k} = \frac{\|\boldsymbol{h}_{i,k}\|^2}{\sigma_\eta^2} \!-\! \frac{1}{\sigma_\eta^4}\boldsymbol{h}_{i,k}^\dag \ddot{\boldsymbol{G}}_{i,k} \ddot{\boldsymbol{V}}_{i,k} \ddot{\boldsymbol{G}}_{i,k}^\dag \boldsymbol{h}_{i,k},
\end{equation}
where
\begin{align}
    \ddot{\boldsymbol{V}}_{i,k} & \triangleq {\left(\boldsymbol{I}_{U-1} + \ddot{\boldsymbol{G}}_{i,k}^\dag \ddot{\boldsymbol{G}}_{i,k}\right)}^{-1} \nonumber \\ 
    &= {\left(\ddot{\boldsymbol{U}}_{i-1}^{-1} - \frac{1}{\sigma_\eta^2}\boldsymbol{u}_k^\dag \boldsymbol{u}_k\right)}^{-1}\nonumber \\
    & = \ddot{\boldsymbol{U}}_{i-1} + \frac{\ddot{\boldsymbol{U}}_{i-1} \boldsymbol{u}_k^\dag \boldsymbol{u}_k \ddot{\boldsymbol{U}}_{i-1}} {\sigma_\eta^2 - \boldsymbol{u}_k \ddot{\boldsymbol{U}}_{i-1} \boldsymbol{u}_k^\dag},
\end{align}
and $\ddot{\boldsymbol{U}}_{i-1} = {(\ddot{\boldsymbol{U}}_{i-2} - \boldsymbol{u}_{k_{i-1}} \boldsymbol{u}_{k_{i-1}}^\dag/\sigma_\eta^2)}^{-1}$ has been computed during the $(i-1)$-th iteration. During the first iteration, $\ddot{\boldsymbol{U}}_{0} = \boldsymbol{U}_N = {(\boldsymbol{I}_{U-1} + \boldsymbol{G}^\dag \boldsymbol{G})}^{-1}$. In this way, we can determine the SINR and select the port successively in a decremental way. 

During each iteration, the port that maximizes the combining SINR of the remaining ports is removed, i.e., $k_i = \arg\max_{k \in \Omega} \gamma_{i,k}$. Subsequently, the index set $\Omega$ is updated by removing $k_i$. This process continues until there are $N_{\rm RF} = N^*$ ports remaining within the index set $\Omega$, which will then constitute the output selected index set $\kappa$. The detailed procedure of this DPS method is summarized in Algorithm~\ref{Alg:DPS}. 

\begin{algorithm} [tbp]
    \caption{Decremental port selection (DPS)}\label{Alg:DPS}
    \begin{algorithmic}[1]
        \Require{$\boldsymbol{g}_{(u,u)}$, $\boldsymbol{\Phi}$, $N$, $N^*$}
        \Ensure{Index set of selected ports $\kappa$}
        \State{Initialization: $\Omega = \{1,2,\dots,N\}$}
        \For{$i = 1$ to $N-N^*$}
            \State{$\omega_{i,k} = \Omega \backslash k,~\forall k \in \Omega$}
            \State{$\boldsymbol{h}_{i,k} = {[\boldsymbol{g}_{(u,u)}]}_{\omega_{i,k}},~\boldsymbol{\Psi}_{i,k} = {[\boldsymbol{\Phi}]}_{\omega_{i,k}, \omega_{i,k}},~ \forall k \in \Omega$}
            \State{Calculate SINR $\gamma_{i,k}$ as~\eqref{Eq:sinrD_n_IPS} or~\eqref{Eq:sinrS_DPS}}
            \State{$k_i = \arg \max_{k \in \Omega} \gamma_{i,k}$}
            \State{Update $\Omega = \Omega \backslash k_i $}
        \EndFor
        \State{$\kappa = \Omega$}
    \end{algorithmic}
\end{algorithm}

We now analyze the complexity of the DPS algorithm. During the $i$-th iteration, we have $\boldsymbol{h}_{i,k}\in \mathbb{C}^{(N-i)\times 1}$. Similar to the IPS method, the number of multiplication operations for direct SINR calculation is given by $C_{\rm D,DPS}^{(i)} = 2{(N-i)}^3 + {(N-i)}^2 + {(N-i)}$, while the number for successive SINR calculations is $C_{\rm S,DPS}^{(i)} = 3U^2 +(N-i-4)U+1$. The total number of SINR calculations during the $i$-th iteration is $N_{\rm DPS}^{(i)} = N-i+1$, resulting in an overall complexity of $\mathcal{O}(\sum_{n=1}^{N-N_{\rm RF}} N_{{\rm DPS}}^{(i)} \times \min\{C_{\rm D,DPS}^{(i)},C_{\rm S,DPS}^{(i)}\})$.

\subsection{Performance Analysis}\label{subsec:UBLB}
Here, we aim to analyze the ASEP of the multi-port slow FAMA system. The ASEP is not available due to the complex correlation among the ports and the intricate combinations involved. Thus, we consider two particular cases and derive the upper and lower bounds of the ASEP.

\subsubsection{Upper Bound on ASEP}
In the scenario where $N = N_{\rm RF} = N^*$, the index set of selected port is given by $\kappa = \Omega$, and the channel vector is represented by $\boldsymbol{h} = \boldsymbol{g}_{(u,u)}$. As such, the SINR of the IRC receiver is expressed as
\begin{equation}
    \mu =  \boldsymbol{w}\boldsymbol{h} = \boldsymbol{g}_{(u,u)}^\dag {\boldsymbol{\Phi}}^{-1}\boldsymbol{g}_{(u,u)}.
\end{equation}
Based on the detector, $\hat{s}_u = \arg\min_{x\in \boldsymbol{\chi}} |\boldsymbol{w}\boldsymbol{r}-x|$ and the $\mu$-dependent instantaneous symbol error probability (SEP) for an $M$-QAM (quadrature amplitude modulation) constellation, $\boldsymbol{\chi}$, is given as \cite{khandelwal2014new}
\begin{multline}\label{eq:insSEP1}
P_E(\mu) = \frac{4(\sqrt{M}-1)}{\sqrt{M}}Q\left(\sqrt{\frac{3\sigma_s^2\mu}{M-1}} \right)\\
 - \frac{4(\sqrt{M}-1)^2}{\sqrt{M}}Q^2\left(\sqrt{\frac{3\sigma_s^2\mu}{M-1}} \right),
\end{multline}
where $|\boldsymbol{\chi}| = M$, and $Q(\cdot)$ denotes the Gaussian Q-function. In the interference-limited scenario, the probability density function (PDF) of $\mu$ is obtained as \cite{shah1998performance}
\begin{equation}\label{Eq:pdf_mu}
    f_\mu (\mu) = \frac{\Gamma(U)}{\Gamma(N^*)\Gamma(U-N^*)}\times \frac{\mu^{N^*-1}}{(1+\mu)^U},~ N^*<U,
\end{equation}
where $\Gamma(\cdot)$ denotes the standard gamma function. The PDF of $\mu$ is independent of the specific form of the covariance matrix $\boldsymbol{\Phi}$.
The statistical expectation of the instantaneous SEP in \eqref{eq:insSEP1} with respect to $\mu$ leads to the expression for the ASEP in this low-resolution scenario. It delineates the upper bound of the ASEP of the system, as specified in \eqref{Eq:ASEP_ub} on the next page.
\begin{figure*}[tb]
\begin{equation}\label{Eq:ASEP_ub}
    \begin{aligned}
        \bar{P}_E^{ub} & = \int_0^\infty P_E(\mu) f_\mu (\mu) d\mu\\
        & = \frac{\Gamma(U)}{\Gamma(N^*)\Gamma(U-N^*)} \times \frac{4(\sqrt{M}-1)}{\sqrt{M}} \times
        \int_0^\infty \left[Q\left(\sqrt{\frac{3\sigma_s^2\mu}{M-1}} \right) - (\sqrt{M}-1)Q^2\left(\sqrt{\frac{3\sigma_s^2\mu}{M-1}} \right) \right] \times \frac{\mu^{N^*-1}}{(1+\mu)^U} d\mu
    \end{aligned}
\end{equation}
\hrulefill
\end{figure*}

\subsubsection{Lower Bound on ASEP}
The SINR of EPS is
\begin{equation}\label{eq:gammatomu}
    \gamma_{\rm EPS} = \max_{i \in \left\{1,\dots,{N\choose N^*}\right\}} \gamma_i \leq \max_{i \in \left\{1,\dots,{N\choose N^*}\right\}} \mu_i,
\end{equation} 
where $\gamma_i$ is the SINR of combination $\omega_i$, and $\mu_i$ represents independent and identically distributed (i.i.d.) random variables with the PDF in \eqref{Eq:pdf_mu}. The variables $\gamma_i$ are interdependent due to the correlation effects and the selection process of the FAS ports. Therefore, the lower bound on ASEP for the EPS can be determined by analyzing the random variable $\hat{\mu} = \max_{i \in \left\{1,\dots,{N\choose N^*}\right\}} \mu_i$. The PDF of $\hat{\mu}$ is given by
\begin{align}
    f_{\hat{\mu}}(\mu) = {N\choose N^*} [F_{\mu_i}(\mu)]^{{N\choose N^*}-1}f_{\mu_i}(\mu),
\end{align}
where $F_{\mu_i}(\mu)$ is the cumulative distribution function (CDF) of $\mu_i$. The CDF is given by \cite{shah1998performance}
\begin{multline}
F_{\mu_i}(\mu) = \frac{\Gamma(U)}{\Gamma(N^*+1)\Gamma(U-N^*)}\\
\times \mu^{N^*}  {}_2F_1\left(U,N^*;N^*+1;-\mu\right),
\end{multline}
where ${}_2F_1(a,b;c;x)$ is the hypergeometric function defined as
\begin{equation}
    {}_2F_1(a,b;c;x) = \sum_{n=0}^{\infty} \frac{(a)_n(b)_n}{(c)_n} \frac{x^n}{n!},
\end{equation}
in which $(\cdot)_n$ is the Pochhammer symbol, defined as $(a)_n = \Gamma(a+n)/\Gamma(a)$. Thus, the lower bound on ASEP can be derived through the statistical expectation of the instantaneous SEP concerning $\hat{\mu}$, as illustrated in \eqref{Eq:ASEP_lb} on the following page.
\begin{figure*}
\begin{multline}\label{Eq:ASEP_lb}
\bar{P}_E^{lb} = N^* \times {N\choose N^*} \times \left(\frac{\Gamma(U)}{\Gamma(N^*+1)\Gamma(U-N^*)}\right)^{{N\choose N^*}} \times \frac{4(\sqrt{M}-1)}{\sqrt{M}}\\
\times \int_0^\infty {\left[  Q\left(\sqrt{\frac{3\sigma_s^2\mu}{M-1}} \right) - (\sqrt{M}-1)Q^2\left(\sqrt{\frac{3\sigma_s^2\mu}{M-1}} \right) \right] \times \frac{\mu^{N^*\times {N\choose N^*}}}{\mu(1+\mu)^U}\times {}_2F_1^{{N\choose N^*}-1} \left(U,N^*;N^*+1;-\mu\right)} d\mu 
\end{multline}
\hrulefill
\end{figure*}

\subsection{Complexity Comparison}
The comparison of complexity among EPS, IPS, and DPS is given in Table~\ref{Tab:complexity}. The complexity of SS for multi-port slow FAMA and CUMA is also included. In the SS method, the calculation of the single-port SINR necessitates $4N$ multiplications, and the final combination requires $(2N_{\rm RF}^3 + N_{\rm RF}^2 + N_{\rm RF})$ multiplications. In CUMA, the multiple ports are combined in the analog domain. Consequently, the multiplication complexity in the digital domain remains relatively low. It involves only the complexity associated with the final combination. 

\begin{table*}
\begin{center}
\caption{Complexity Comparison of Port Selection Methods for Multi-port \emph{s}-FAMA}\label{Tab:complexity}
\resizebox{.95\linewidth}{!}{
\begin{tabular}{l|c|c|c}
    \hline
    Method & $C_{\rm D}^{(i)}$ for $i$-th iteration & $C_{\rm S}^{(i)}$ for $i$-th iteration & Complexity\\ \hline\hline
    EPS & $2N_{\rm RF}^3+N_{\rm RF}^2+N_{\rm RF}$  & $(2N_{\rm RF}+1)(U^2-U)+N_{\rm RF}$ & $\mathcal{O}\left(C({N},{N_{\rm RF}}) \times \min\{C_{\rm D},C_{\rm S}\}\right)$ \\ \hline
    IPS & $2i^3+i^2+i$  & $3U^2 +(i-4)U +1$ & $\mathcal{O}\left(\sum_{n=1}^{N_{\rm RF}} (N-n+1) \times \min\{C_{\rm D},C_{\rm S}\}\right)$ \\ \hline
    DPS & $2{(N-i)}^3 + {(N-i)}^2 + {(N-i)}$ & $3U^2 +(N-i-4)U+1$ & $\mathcal{O}\left(\sum_{n=1}^{N-N_{\rm RF}} (N-n+1) \times \min\{C_{\rm D},C_{\rm S}\}\right)$ \\ \hline
    SS in \cite{hong2025Downlink} & -- & -- & $\mathcal{O}\left(2N_{RF}^3+N_{RF}^2+N_{RF}+4N \right)$\\ \hline
    CUMA in \cite{Wong2024cuma} & -- & -- & $\mathcal{O}\left(2N_{RF}^3+N_{RF}^2+N_{RF}\right)$\\\hline    
\end{tabular}}
\end{center}
\end{table*}

\begin{figure}
\begin{center}
\subfigure[$N = 15$]{\includegraphics[width=\linewidth]{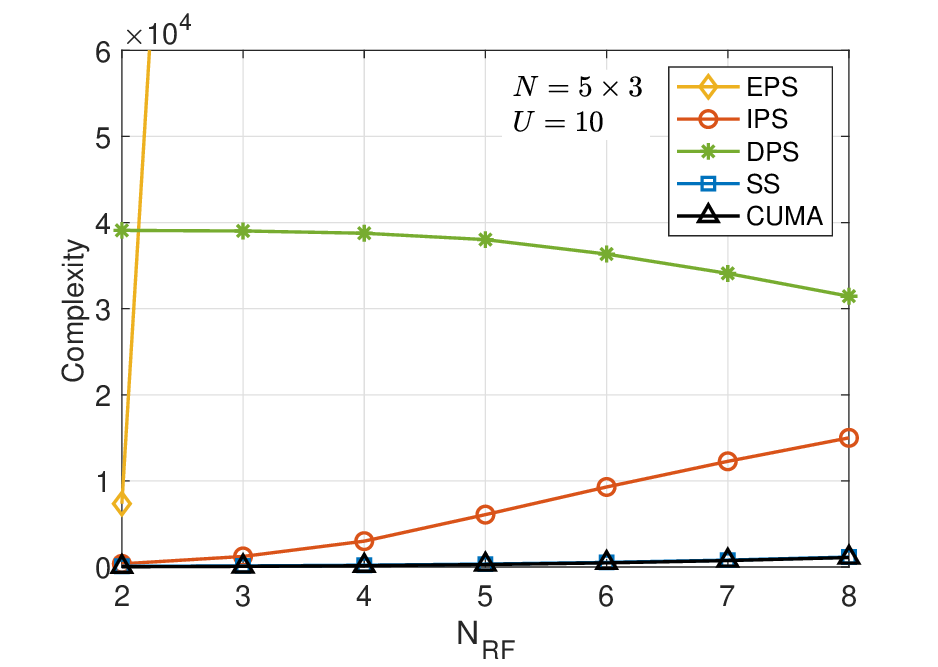}\label{SubFig:CompN15}}
\subfigure[$N = 450$]{\includegraphics[width=\linewidth]{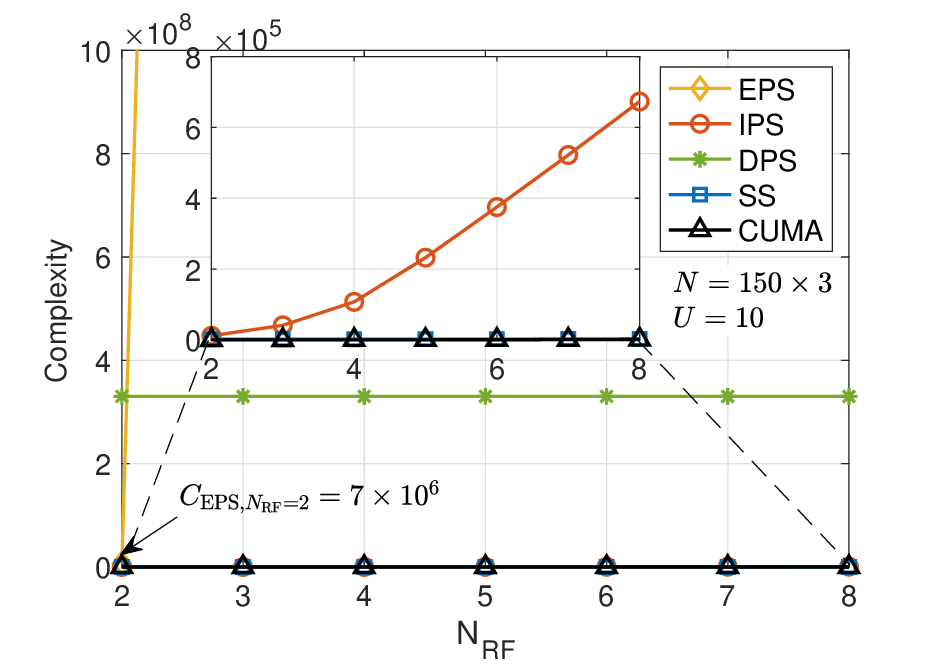}\label{SubFig:CompN450}}
\caption{Numerial complexity comparison for various port selection methods, with $U=4$, (a) $N=15$, and (b) $N = 450$.}\label{Fig:Complexity}
\vspace{-2mm}
\end{center}
\end{figure}

The numerical complexity comparison for $U=10$ is presented in Fig.~\ref{Fig:Complexity}, where Fig.~\ref{SubFig:CompN15} assumes a relatively sparse case with $N = 15$ and Fig.~\ref{SubFig:CompN450} considers a compact case with a large number of ports. Among the proposed methods, the IPS method exhibits the relatively lowest complexity. It is slightly higher than that of the SS method and CUMA. Furthermore, the complexity of the IPS method increases with the number of RF chains, $N_{\rm RF}$. The complexity of the EPS method increases rapidly with $N_{\rm RF}$, indicating that it is only practical in scenarios with low $N_{\rm RF}$, such as in the case with $N_{\rm RF} = 2$. The complexity of the DPS method exceeds that of EPS when $N_{\rm RF} = 2$. It decreases as $N_{\rm RF}$ increases, particularly when $N_{\rm RF} \to N$, where the complexity becomes more manageable. The complexity of DPS is deemed acceptable only when $N$ is small. However, as illustrated in Fig.~\ref{SubFig:CompN450} with a larger $N$, DPS becomes excessively complex. It can be concluded that DPS is practical in scenarios characterized by low $N$ and relatively high $N_{\rm RF}$, such as the case with $N = 15$ and $N_{\rm RF}=8$. 

In summary, considering the complexity, EPS is only suitable for $N_{\rm RF} = 2$, IPS works well in the case of $N_{\rm RF} \ll N$, and DPS is desirable for $N_{\rm RF} \to N$.

\subsection{Estimated Covariance Matrix}\label{subsec:ECM}
In practice, it is generally infeasible to accurately estimate the channel vectors for interferers, i.e., the unknown interference channel matrix $\boldsymbol{G}$. Assuming the channel vector $\boldsymbol{g}_{(u,u)}$ is known at the receiver for the $u$-th UT, the covariance matrix can be estimated  during the training periods as
\begin{equation}
\widetilde{\boldsymbol{\Phi}} = \frac{1}{T} \sum_{t=1}^{T} {\boldsymbol{r}}[t] {\boldsymbol{r}}^\dag [t] - \boldsymbol{g}_{(u,u)} \boldsymbol{g}_{(u,u)}^\dag,
\end{equation}
where $T$ denotes the number of received signal vectors. 

Assuming that $\sigma_\eta^2$ is known at the receiver, $\boldsymbol{G}$ can be estimated using the singular value decomposition as 
\begin{equation}
\widetilde{\boldsymbol{\Phi}} - \sigma^2_\eta \boldsymbol{I}_N = \boldsymbol{U}\boldsymbol{\Lambda}\boldsymbol{U}^\dag = \boldsymbol{U}\boldsymbol{\Lambda}^{\frac{1}{2}}\boldsymbol{\Lambda}^{\frac{1}{2}}\boldsymbol{U}^\dag = \widetilde{\boldsymbol{G}} \widetilde{\boldsymbol{G}}^\dag,
\end{equation}
where $\widetilde{\boldsymbol{G}}$ is the estimated channel matrix comprising $(U-1)$ non-zero columns of $(\boldsymbol{U}\boldsymbol{\Lambda}^{{1}/{2}})$. With the above estimation of the covariance matrix, the proposed algorithms outlined in this section may be reused for port selection purposes.

\section{Performance Evaluation}\label{sec:Perf}
This section presents simulation results aimed at evaluating the performance of the port selection methods. The parameters used in the simulations are detailed in Table~\ref{Tab:SimPara}, reflecting typical scenarios. The actual dimensions of the 2D-FAS at each UT are set to $15~{\rm cm}\times 8~{\rm cm}$, approximating the size of a typical mobile handset. The case at $6~{\rm GHz}$ frequency corresponds to the 5G Mid-band with a bandwidth of $10~{\rm MHz}$, whereas the $26~{\rm GHz}$ case pertains to the millimeter-wave band with a bandwidth of $50~{\rm MHz}$. The channel parameters $(K,N_p)$ for the finite-scattering channel model are selected carefully on reflecting the characteristics of the respective frequency. We focus on the practical case where the channel vectors for interferers, $\boldsymbol{g}_{(\tilde{u},u)}$ ($\forall \tilde{u} \neq u$), are unknown, and reception is based on the estimated covariance matrix, as in Section \ref{subsec:ECM}.

\begin{table}[t]
\begin{center}
    \vspace{-2mm}
    \caption{Simulation Parameters}\label{Tab:SimPara}
\resizebox{.95\columnwidth}{!}{
    \begin{tabular}{l|c|c|c|c}
        \hline
        \textbf{Parameter}                  & \multicolumn{4}{l}{\textbf{Value}} \\ \hline\hline
        Carrier frequency, $f$  & \multicolumn{2}{c|}{$6 ~{\rm GHz}$}   & \multicolumn{2}{c}{$26 ~{\rm GHz}$} \\ \hline
        Wavelength, $\lambda$   & \multicolumn{2}{c|}{$5 ~{\rm cm}$}    & \multicolumn{2}{c}{$1.15 ~{\rm cm}$} \\ \hline
        FAS size, $W$           & \multicolumn{2}{c|}{$3\lambda \times 1.6 \lambda$}    & \multicolumn{2}{c}{$13\lambda \times 7\lambda$} \\ \hline
        Number of ports, $N$    & $5\times 3$   & $150\times 3$ & $25\times 13$ & $64\times 13$ \\ \hline
        Bandwidth, $B$   & \multicolumn{2}{c|}{$10~{\rm MHz}$}    & \multicolumn{2}{c}{$50~{\rm MHz}$} \\ \hline
        Subframe duration & \multicolumn{2}{c|}{$1~{\rm ms}$}    & \multicolumn{2}{c}{$1/2^2~{\rm ms}$} \\ \hline 
        Subcarrier spacing, $\Delta f$   & \multicolumn{2}{c|}{$15~{\rm kHz}$}    & \multicolumn{2}{c}{$60~{\rm kHz}$} \\ \hline
        Number of PRB, $N_{RB}$   & \multicolumn{2}{c|}{$52$}    & \multicolumn{2}{c}{$66$} \\ \hline
        Cyclic prefix (CP)  & \multicolumn{4}{c}{Normal}  \\ \hline
        \makecell[l]{Number of resource \\ elements in a PRB, $N'_\text{RE}$} & \multicolumn{4}{c}{$156$} \\ \hline
        FFT Size, $N_\text{fft}$ & \multicolumn{4}{c}{$1024$} \\ \hline
        $[\rho,N_{\max}]$ for CUMA & $[0.2,3]$ & \multicolumn{3}{c}{$[0.2,80]$}\\ \hline
        Demapping scheme    & \multicolumn{4}{c}{Max-Log-MAP} \\ \hline
        Decoding scheme     & \multicolumn{4}{c}{Min-Sum {with $50$ iterations}}      \\ \hline \hline
        \multicolumn{5}{c}{Finite-scattering Channels} \\ \hline
        Rice factor, $K$    & \multicolumn{2}{c|}{$0$}  & \multicolumn{2}{c}{$7$} \\ \hline
        No. of NLOS paths, $N_p$    & \multicolumn{2}{c|}{$50$}  & \multicolumn{2}{c}{$2$} \\ \hline\hline
        \multicolumn{5}{c}{CDL channels} \\ \hline
        Type & \multicolumn{2}{c|}{CDL-C} & \multicolumn{2}{c}{CDL-E}\\ \hline
        Delay spread &  \multicolumn{2}{c|}{$30 ~{\rm ns}$} & \multicolumn{2}{c}{$20 ~{\rm ns}$} \\ \hline
    \end{tabular}}
\end{center}
\end{table}

We first study the uncoded symbol error rate (SER) performance of the proposed methods. Subsequently, we utilize the 5G NR modulation and coding schemes (MCSs) for physical downlink shared channel (PDSCH) \cite{38214} and present the block error rate (BLER) performance from link-level simulations. With these results, we derive and exhibit the multiplexing gains of the system employing various port selection methods. Finally, the system is extended to a wideband configuration, wherein the proposed methods are evaluated in CDL channels \cite{38901}. The parameters pf CDL channel model are selected to reflect the characteristics of the frequency.

\vspace{-2mm}
\subsection{Uncoded SER Performance}

\begin{figure}
\centering
\subfigure[$f=6~{\rm GHz}$ (sparse)]{\includegraphics[width=\linewidth]{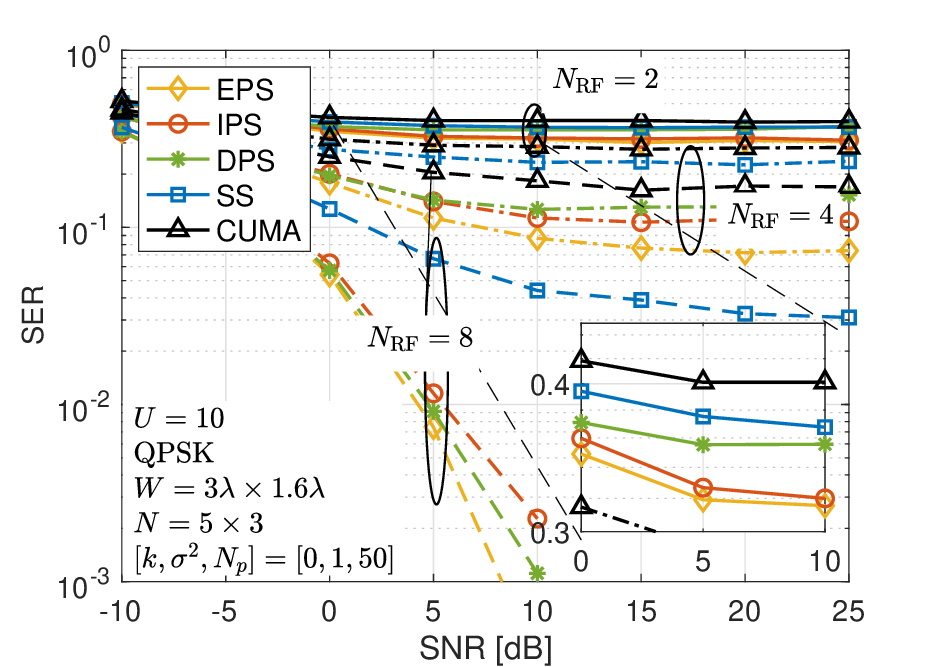}\label{SubFig:SERvsSNR_N5x3}}
\subfigure[$f=6~{\rm GHz}$ (compact)]{\includegraphics[width=\linewidth]{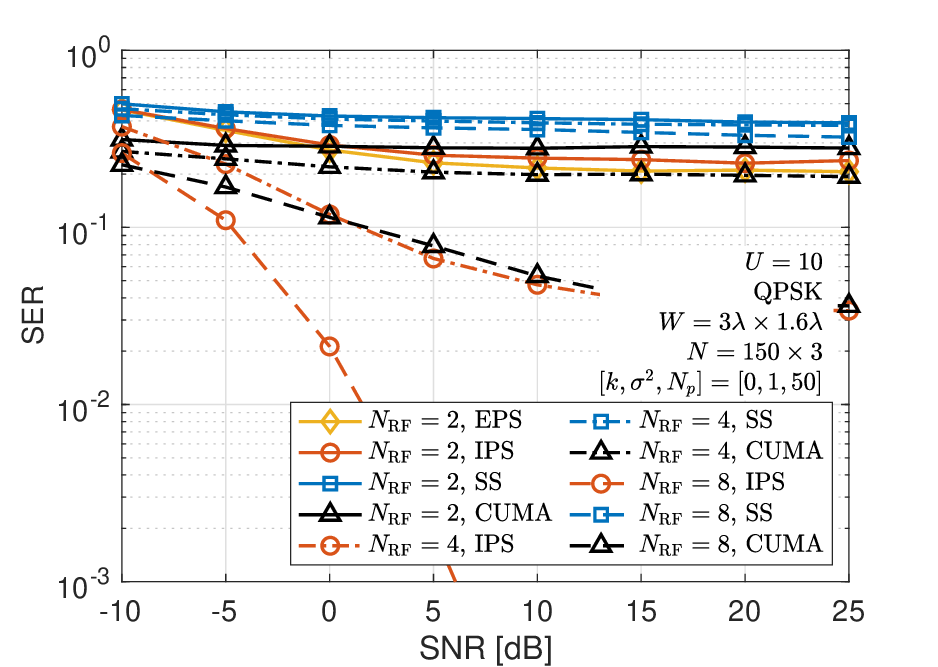}\label{SubFig:SERvsSNR_N150x3}}
\caption{SER performance against transmit SNR for different schemes under finite-scattering.}\label{fig:SERvsSNR}
\vspace{-2mm}
\end{figure}

The results in Fig.~\ref{fig:SERvsSNR} are presented for the SER of slow FAMA with the proposed port selection methods against the average transmit SNR, $\sigma^2\sigma_s^2/\sigma_\eta^2$. The carrier frequency is $6~{\rm GHz}$ and quadrature phase shift keying (QPSK) is employed in the simulations. In addition, the study emphasizes large-$U$ massive connectivity scenarios, with the number of UTs set to $U=10$. Benchmark results for FAMA using the SS method \cite{hong2025Downlink} and CUMA \cite{Wong2024cuma} are provided for comparative purposes. Fig.~\ref{SubFig:SERvsSNR_N5x3} illustrates a sparse FAS case, where $N = 5\times 3$ ports are uniformly distributed within the physical area of $W = 3\lambda \times 1.6\lambda$. In this case, all three port selection methods are simulated. The parameters of CUMA are set as $\rho = 0.2$, and $N_{\max} = 3$. Conversely, in the compact configuration with $N = 150\times 3$, as shown in Fig.~\ref{SubFig:SERvsSNR_N150x3}, the EPS and IPS methods are employed, with EPS being adopted exclusively when $N_{\rm RF} = 2$. For this case, the parameters of CUMA are chosen as $\rho = 0.2$, and $N_{\max} = 80$. The solid curves represent the results with $N_{\rm RF} = 2$, the dash-dot curves depict results for the case $N_{\rm RF} = 4$, and the dashed curves correspond to $N_{\rm RF} = 8$. As expected, the proposed port selection techniques surpass the SS method, and slow FAMA with the proposed methods even attains superior performance to CUMA. 

Analyzing the results in Fig.~\ref{SubFig:SERvsSNR_N5x3}, it is evident that slow FAMA with DPS surpasses slow FAMA with SS and CUMA, aligning with the results reported in \cite{coma2024slow}, where GEPort is characterized as a variant of DPS with an optimized combining vector $\boldsymbol{w}$. When $N_{\rm RF}$ is small, such as when $N_{\rm RF} = 2$ or $4$, IPS demonstrates superior performance compared to DPS, because the decremental strategy might erroneously discard ports with better performance during the iteration process of DPS. On the contrary, when $N_{\rm RF}$ is relatively large, such as $N_{\rm RF} = 8$, DPS outperforms IPS. Furthermore, EPS surpasses both IPS and DPS but entails high complexity. Note that in this sparse case, CUMA is even inferior to slow FAMA with SS, indicating that CUMA may not be suitable for cases involving a small number of FAS ports at lower frequencies. 

In Fig.~\ref{SubFig:SERvsSNR_N150x3}, the performance of slow FAMA with SS is poor because of inaccuracies in single-port SINR assessment. Conversely, IPS and EPS demonstrate performance improvements. Moreover, it is noteworthy to observe the presense of an error floor attributed to the large-$U$ condition. In scenarios of massive connectivity, the primary source of contamination is interference rather than additive noise. Henceforth, focus is directed towards interference-limited scenarios where noise effects are negligible, Specifically, the average transmit SNR, $\sigma^2\sigma_s^2/\sigma_\eta^2$, will be set to a high value, such as $35~{\rm dB}$.

\begin{figure}
\subfigure[$f=6~{\rm GHz}$, $N_{\rm RF} = 2$]{\includegraphics[width = \linewidth]{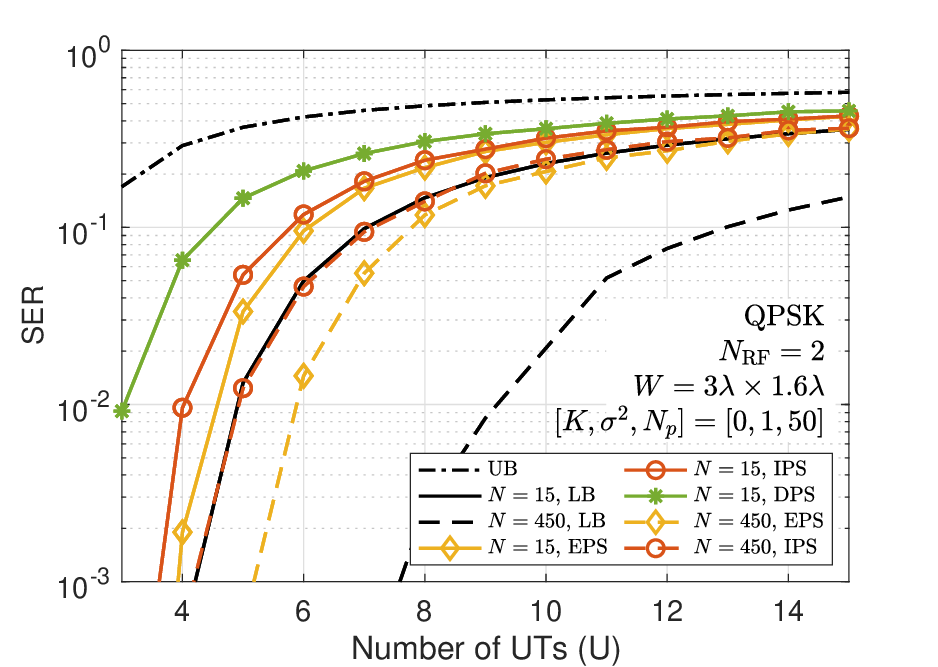}}
\subfigure[$f=6~{\rm GHz}$, $N_{\rm RF} = 4$]{\includegraphics[width = \linewidth]{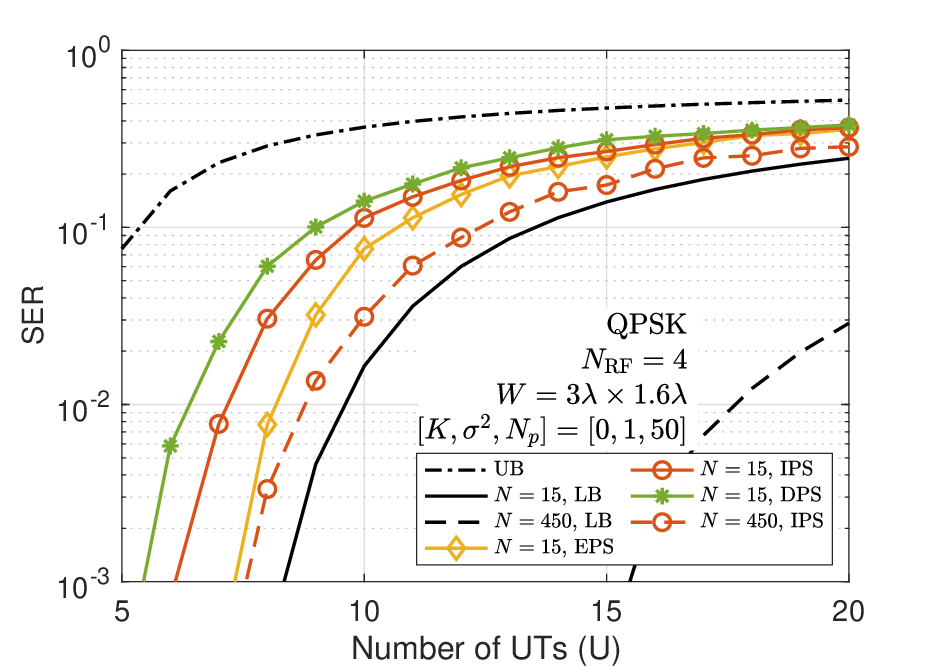}}
\caption{Upper and lower bounds on ASEP against number of UTs. Empirical SER results under finite-scattering are presented for comparison.}\label{fig:PEvsU}
\vspace{-2mm}
\end{figure}

Fig.~\ref{fig:PEvsU} compares the empirical SER results with the upper and lower bounds derived in Section \ref{subsec:UBLB}. It can be seen that the empirical curves are bounded between these upper and lower bounds. In the sparse configuration when the number of FAS ports is $N = 15$, the lower bound is tight, and the SER curve of EPS approaches the lower bound. However, in the high-resolution scenario ($N = 450$), the lower bound appears overestimated. This is because, as indicated in \eqref{eq:gammatomu}, the lower bound can be considered as the ASEP of an idealized system with no correlation among the combining SINR. For a genuine FAMA system, when $W$ is fixed and $N$ increases, the distance between two adjacent ports decreases, and the correlation between these ports intensifies, resulting in an increase in correlation among $\gamma_i$ of different combinations. Thus, the empirical results diverge from the lower bound.

\begin{figure*}
\centering
\subfigure[$f=6~{\rm GHz}$ (sparse)]{\includegraphics[width=0.48\linewidth]{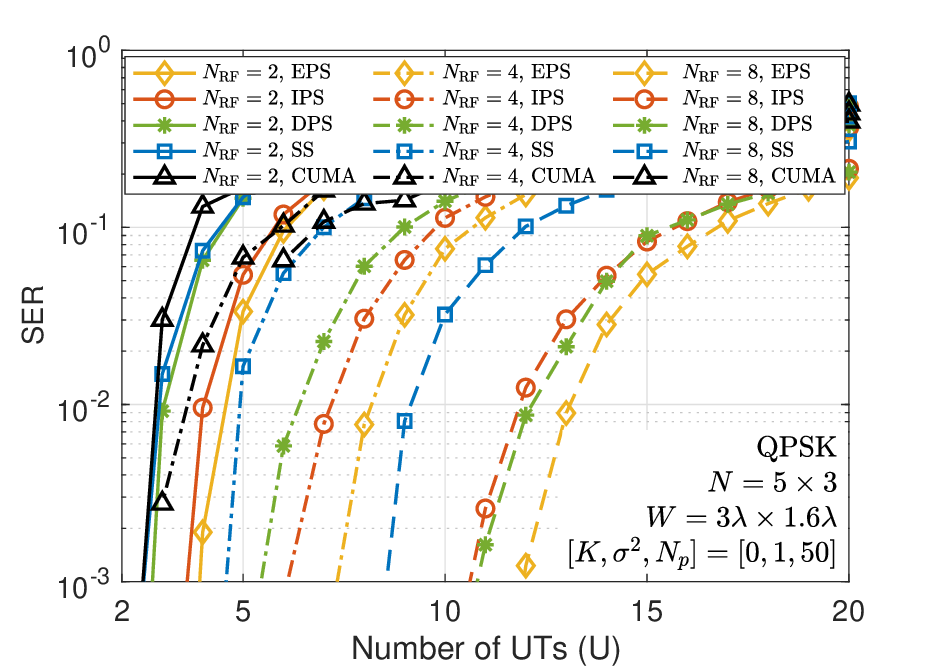}\label{SubFig:SERvsU_N5x3}}
\subfigure[$f=6~{\rm GHz}$ (compact)]{\includegraphics[width=0.48\linewidth]{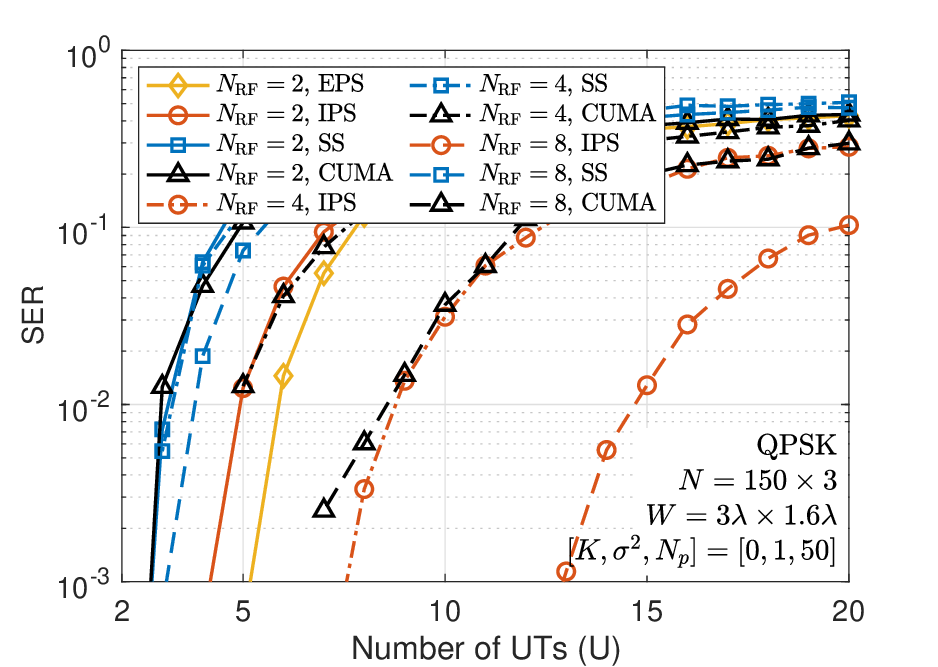}\label{SubFig:SERvsU_N150x3}} \vspace{-3mm}\\
\subfigure[$f=26~{\rm GHz}$ (sparse)]{\includegraphics[width=0.48\linewidth]{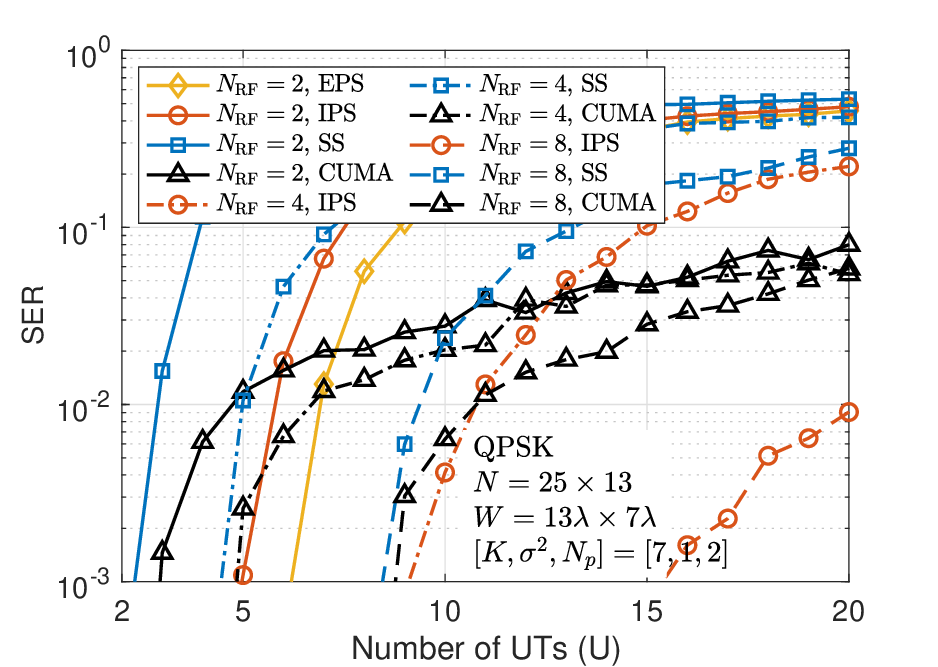}\label{SubFig:SERvsU_N25x13}}
\subfigure[$f=26~{\rm GHz}$ (compact)]{\includegraphics[width=0.48\linewidth]{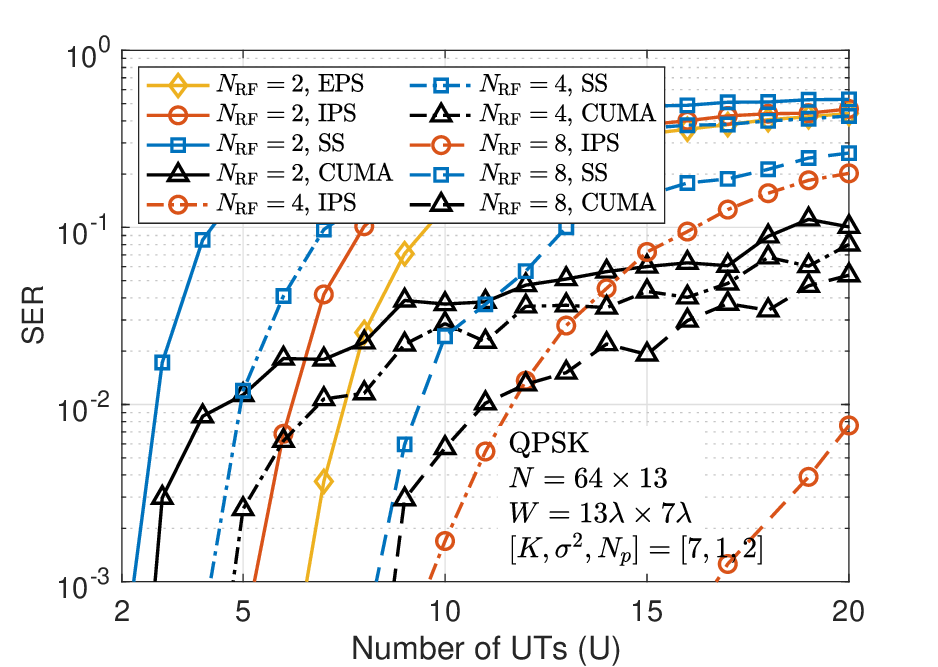}\label{SubFig:SERvsU_N64x13}}
\caption{SER performance against the number of UTs for different schemes under interference limited finite-scattering channels.}\label{fig:SERvsU}\vspace{-2mm}
\end{figure*}

Additional SER performance results at various frequencies are presented in Fig.~\ref{fig:SERvsU}. Figs.~\ref{SubFig:SERvsU_N5x3} and \ref{SubFig:SERvsU_N150x3} show the results at $6~{\rm GHz}$, where Fig.~\ref{SubFig:SERvsU_N5x3} pertains to a sparse configuration with $N=5\times 3$ and Fig.~\ref{SubFig:SERvsU_N150x3} corresponds to a compact configuration with $N = 150\times 3$. In the sparse scenario, CUMA does not perform well because the normalized size of FAS is too small and the number of antenna ports is limited, thereby hindering CUMA from leveraging its advantages in analog combining. Conversely, slow FAMA reveals robust performance under this condition, with improvements linking to the increase in the number of RF chains. In slow FAMA, the three proposed port selection methods (EPS, IPS, and DPS) outperform the SS method. Among these, EPS consistently yields the superior performance, at the cost of high complexity. When $N_{\rm RF}$ is low, IPS outperforms DPS, and vice versa. In the compact scenario depicted in Fig.~\ref{SubFig:SERvsU_N150x3}, the two benchmarks, CUMA and slow FAMA with SS, exhibit comparable SER performance when $N_{\rm RF} = 2$, attributable to strongly correlated received signals at this constrained FAS size. But as $N_{\rm RF}$ increases, CUMA shows superior performance to slow FAMA with SS, with the gains of the latter being marginal. This slight performance gain in the compact case indicates the inaccuracy of the SS strategy for highly compact FAS configurations. Conversely, slow FAMA with IPS performs effectively and even surpasses CUMA. The number of UTs supported by this strategy is approximately $15$ when $N_{\rm RF}=8$, $9$ when $N_{\rm RF}=4$, and $5$ when $N_{\rm RF}=2$, for QPSK at a target of ${\rm SER} = 0.01$.

Figs.~\ref{SubFig:SERvsU_N25x13} and \ref{SubFig:SERvsU_N64x13} illustrate the SER results at $26~{\rm GHz}$. In Fig.~\ref{SubFig:SERvsU_N25x13}, it is assumed that the FAS at the receiver has ports separated by a minimum of half a wavelength, whereas Fig.~\ref{SubFig:SERvsU_N64x13} considers a more compact configuration. First, the results in Fig.~\ref{SubFig:SERvsU_N25x13} reveal that the trend across various port selection methods is consistent with observations at the lower frequency. Specifically, IPS outperforms SS, and EPS exhibits the best performance when $N_{\rm RF} = 2$. Notably, the CUMA SER curve remains relatively flat. When $U$ is large, CUMA attains a marginally lower SER than slow FAMA with IPS or EPS. However, when $U$ is small, slow FAMA with IPS or EPS reaches a waterfall performance more rapidly. This suggests that slow FAMA with the proposed port selection methods may deliver superior performance under constrained connectivity, but CUMA might achieve a higher multiplexing gain when employing highly robust MCS. In Fig.~\ref{SubFig:SERvsU_N64x13}, the results closely resemble those in Fig.~\ref{SubFig:SERvsU_N25x13}, indicating that denser packing of antenna ports at the UT at high frequency appears to be neither advantageous nor detrimental.

\vspace{-2mm}
\subsection{Coded BLER Performance}\label{subsec:bler_perf}

\begin{figure}
\centering
\subfigure[$f=6~{\rm GHz}$ (compact)]{\includegraphics[width = \linewidth]{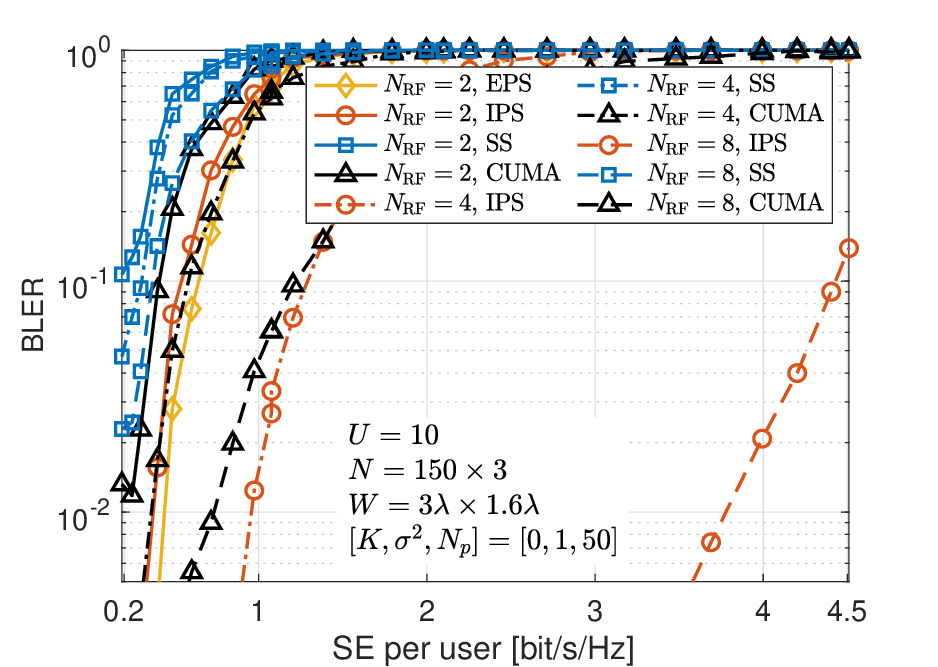}\label{subfig:BLERvsSE6GHz}}
\subfigure[$f=26~{\rm GHz}$ (compact)]{\includegraphics[width = \linewidth]{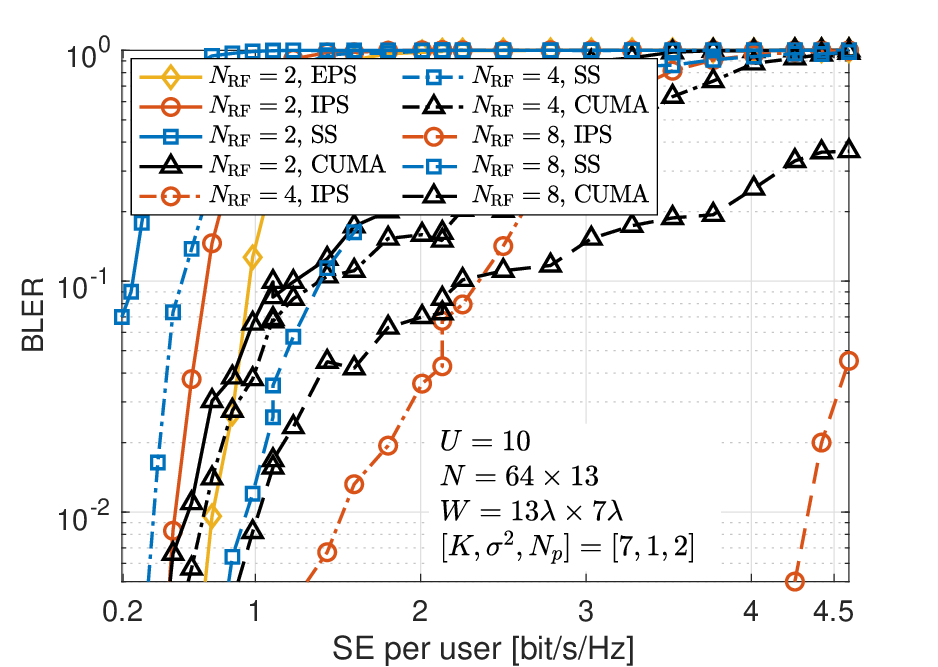}\label{subfig:BLERvsSE26GHz}}
\caption{BLER performance against SE under finite-scattering.}\label{fig:BLERvsSE}
\vspace{-2mm}
\end{figure}

In Fig.~\ref{fig:BLERvsSE}, we illustrate the BLER results relative to the spectral efficiency (SE) per user, with $U = 10$. Two compact scenarios are considered, where Fig.~\ref{subfig:BLERvsSE6GHz} depicts the results at a lower frequency of $f = 6~{\rm GHZ}$, and Fig.~\ref{subfig:BLERvsSE26GHz} pertains to a high frequency of $f = 26~{\rm GHz}$. The MCSs employed range from MCS 0 to 28 \cite[Table 5.1.3.1-1]{38214}, with SE varying from $0.2$ to $4.5$ bit/s/Hz. The SE per UT is calculated as
\begin{equation}
{\rm SE} = {\rm TBS}/(B\times T_{\rm subframe}),
\end{equation}
where $\rm TBS$ represents the transmit block size determined by the target code rate and modulation order specified in \cite[Table 5.1.3.1-1]{38214}, $B$ denotes the bandwidth, and $T_{\rm subframe}$ is the duration of a subframe.

As expected, the system with lower SE can achieve lower BLER performance. For an equivalent BLER, slow FAMA with the proposed port selection methods can support higher SE in transmission compared to SS. Specifically, when $N_{\rm RF} = 2$, EPS outperforms IPS, with IPS superior to SS. When $N_{\rm RF} = 4$ or $8$, the simulations for EPS were not conducted due to the extremely high complexity, but IPS continues to outperform SS. Moreover, the performance improves with increasing $N_{\rm RF}$. For example, in the low frequency band, under the criterion of ${\rm BLER} = 0.01$, the per-user SE of IPS can reach approximately $0.4$ bit/s/Hz with $N_{\rm RF} = 2$ RF chains, which can be increased to $1.0$ bit/s/Hz with $N_{\rm RF} = 4$ and $3.8$ bit/s/Hz with $N_{\rm RF} = 8$. Nevertheless, this enhancement is relatively limited in the case of SS or CUMA. Compared to CUMA, we observe from Fig.~\ref{subfig:BLERvsSE6GHz} that slow FAMA with the proposed methods consistently exhibits advantages in the low frequency band. However, in the high frequency band depicted in Fig.~\ref{subfig:BLERvsSE26GHz}, CUMA yields marginally lower BLER in cases with higher SE. Conversely, in the low SE region, slow FAMA with the proposed IPS method attains comparable or superior BLER performance relative to CUMA.

\vspace{-2mm}
\subsection{Multiplexing Gain}
Another important key performance indicator is the multiplexing gain, which can be calculated as 
\begin{equation}
M = \begin{cases}
U\times (1-{\rm BLER}), & \text{if } {\rm BLER} < {\rm BLER}_{\rm out},\\
0,  & \text{otherwise},
\end{cases}
\end{equation}
where ${\rm BLER}_{\rm out} = 0.1$ is chosen as the out-of-sync BLER. When BLER exceeds ${\rm BLER}_{\rm out}$, reliable data transmission becomes impossible, and the receiver might request retransmissions to enhance data block reception.

\begin{figure*}
\centering
\subfigure[$f = 6~{\rm GHz}$ (Compact), $N_{\rm RF} = 2$]{\includegraphics[width = 0.32\linewidth]{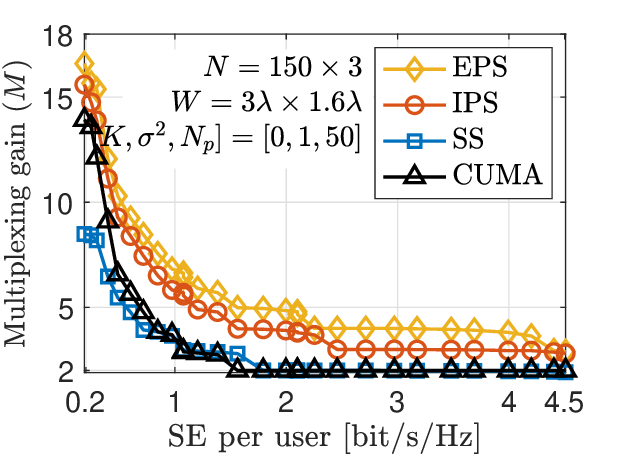}\label{subfig:MGvsSE_BFChan_6GHz_NRF2}}
\subfigure[$f = 6~{\rm GHz}$ (Compact), $N_{\rm RF} = 4$]{\includegraphics[width = 0.32\linewidth]{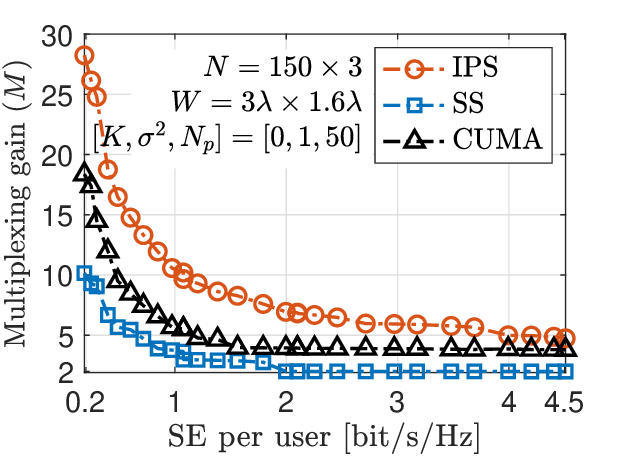}\label{subfig:MGvsSE_BFChan_6GHz_NRF4}}
\subfigure[$f = 6~{\rm GHz}$ (Compact), $N_{\rm RF} = 8$]{\includegraphics[width = 0.32\linewidth]{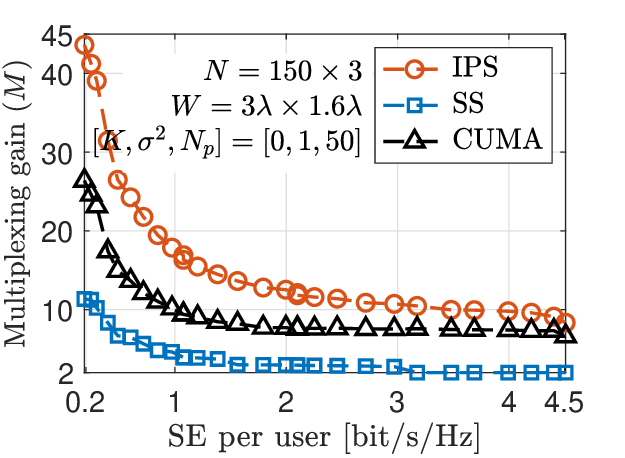}\label{subfig:MGvsSE_BFChan_6GHz_NRF8}}\\
\subfigure[$f = 26~{\rm GHz}$ (Compact), $N_{\rm RF} = 2$]{\includegraphics[width = 0.32\linewidth]{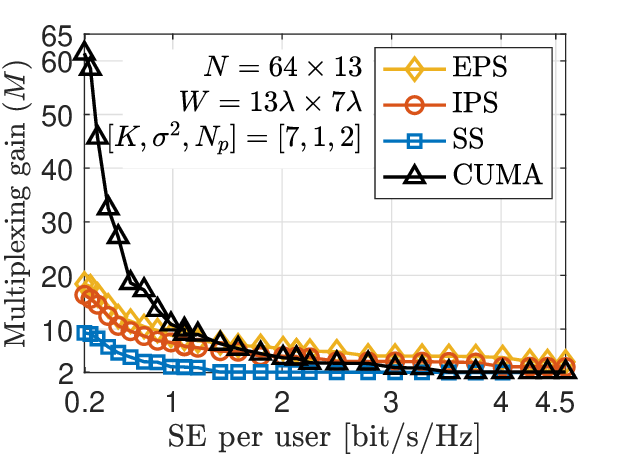}\label{subfig:MGvsSE_BFChan_26GHz_NRF2}}
\subfigure[$f = 26~{\rm GHz}$ (Compact), $N_{\rm RF} = 4$]{\includegraphics[width = 0.32\linewidth]{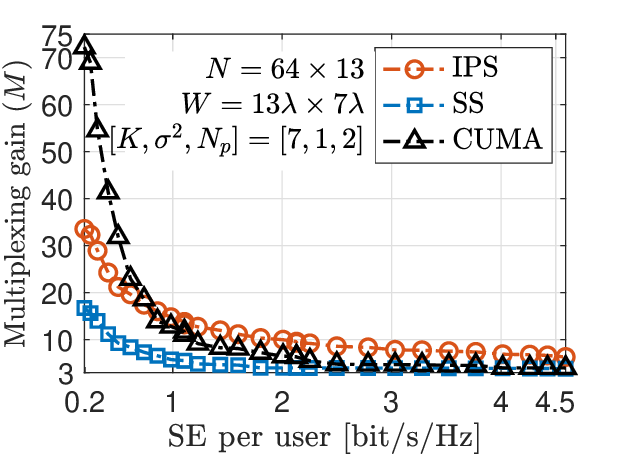}\label{subfig:MGvsSE_BFChan_26GHz_NRF4}}
\subfigure[$f = 26~{\rm GHz}$ (Compact), $N_{\rm RF} = 8$]{\includegraphics[width = 0.32\linewidth]{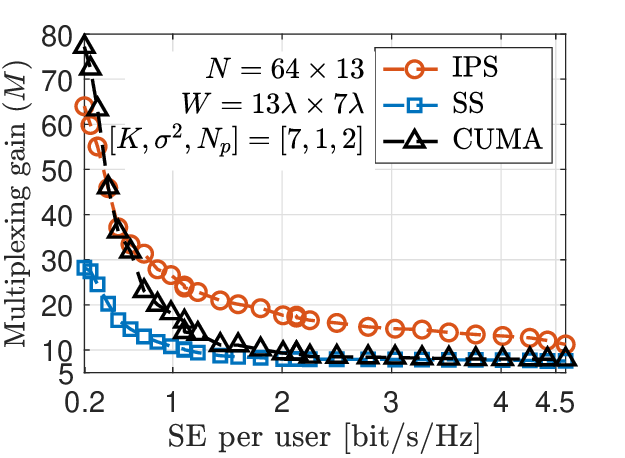}\label{subfig:MGvsSE_BFChan_26GHz_NRF8}}
\caption{Multiplexing gain with different MCS under finite-scattering channels.}\label{fig:MGvsSE_BFChan}
\end{figure*}

Fig.~\ref{fig:MGvsSE_BFChan} illustrates the multiplexing gain results versus SE per UT. The analysis considers only two compact configurations, as the compact setup optimally exploits the capabilities of FAS. The multiplexing gain increases as SE per UT decreases. In other words, the FAMA system is capable of delivering higher multiplexing gains with a robust low-rate MCS. Furthermore, an increase in $N_{\rm RF}$ generally enhances the multiplexing gain across all schemes. The improvement is more noticeable for the proposed IPS method compared to CUMA and SS. The results of EPS are only presented when $N_{\rm RF} = 2$, owing to the prohibitive complexity associated with EPS when $N_{\rm RF}>2$. Moreover, when $N_{\rm RF} = 2$, the performance gain of EPS over IPS is relatively marginal, indicating that IPS might be served as a pragmatic alternative when considering the trade-off between performance and complexity.

At the frequency band of $6~{\rm GHz}$, slow FAMA with the proposed EPS and IPS methods outperform both CUMA and slow FAMA with SS. The gain becomes more conspicuous with an increase in $N_{\rm RF}$. The system is capable of achieving a multiplexing gain exceeding $40$ UTs when $N_{\rm RF} =8$, utilizing MCS 0 or 1. Importantly, even with a limited number of RF chains at the receiver, the multiplexing gain can readily surpass $10$.
At the frequency band of $26~{\rm GHz}$, slow FAMA with IPS or EPS outperforms that with SS. Relative to CUMA, CUMA outperforms the other schemes in the low SE region, but slow FAMA with IPS or EPS offers a higher multiplexing gain in the high SE region. The gap between slow FAMA with IPS and CUMA diminishes as $N_{\rm RF}$ increases. For instance, when $N_{\rm RF} = 2$, the multiplexing gain of CUMA exceeds $60$, whereas that of slow FAMA with IPS is approximately $20$ under MCS 0. As $N_{\rm RF}$ increases to $4$, the multiplexing gains of CUMA and slow FAMA with IPS increase to $72$ and $34$, respectively. When $N_{\rm RF} = 8$, their multiplexing gains become very similar, $77$ for CUMA and $64$ for FAMA with IPS.

\begin{figure*}
\centering
\subfigure[$f = 6~{\rm GHz}$ (Compact), $N_{\rm RF} = 2$]{\includegraphics[width = 0.32\linewidth]{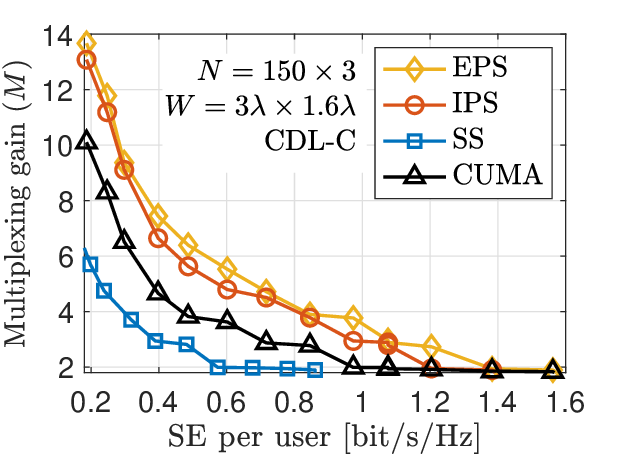}\label{subfig:MGvsSE_CDLChan_6GHz_NRF2}}
\subfigure[$f = 6~{\rm GHz}$ (Compact), $N_{\rm RF} = 4$]{\includegraphics[width = 0.32\linewidth]{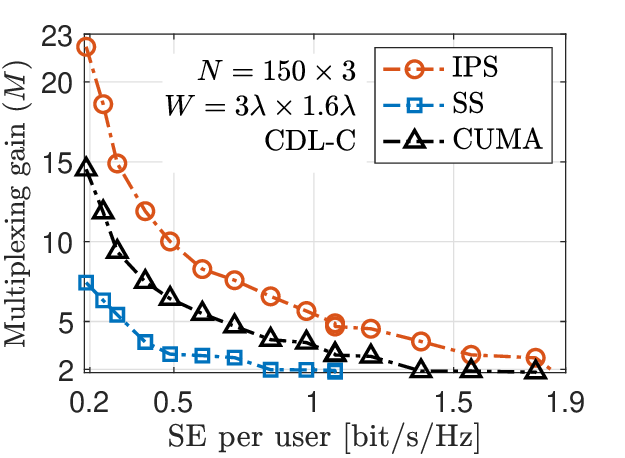}\label{subfig:MGvsSE_CDLChan_6GHz_NRF4}}
\subfigure[$f = 6~{\rm GHz}$ (Compact), $N_{\rm RF} = 8$]{\includegraphics[width = 0.32\linewidth]{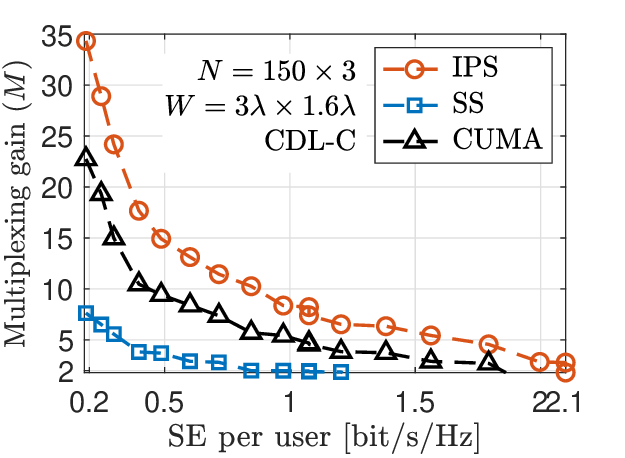}\label{subfig:MGvsSE_CDLChan_6GHz_NRF8}}\\
\subfigure[$f = 26~{\rm GHz}$ (Compact), $N_{\rm RF} = 2$]{\includegraphics[width = 0.32\linewidth]{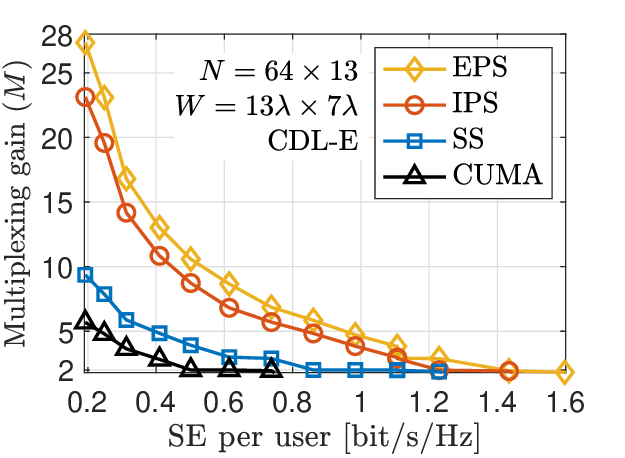}\label{subfig:MGvsSE_CDLChan_26GHz_NRF2}}
\subfigure[$f = 26~{\rm GHz}$ (Compact), $N_{\rm RF} = 4$]{\includegraphics[width = 0.32\linewidth]{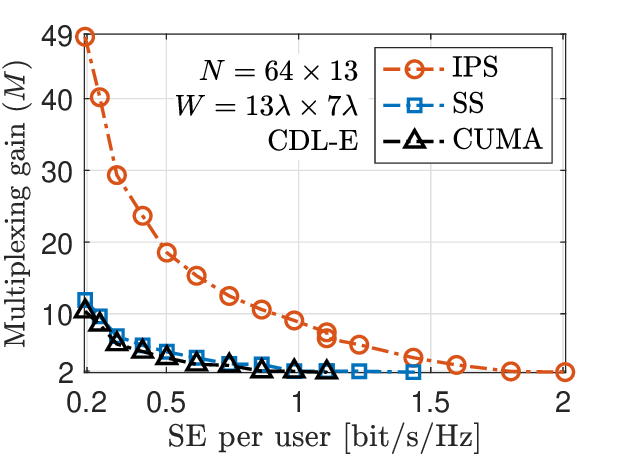}\label{subfig:MGvsSE_CDLChan_26GHz_NRF4}}
\subfigure[$f = 26~{\rm GHz}$ (Compact), $N_{\rm RF} = 8$]{\includegraphics[width = 0.32\linewidth]{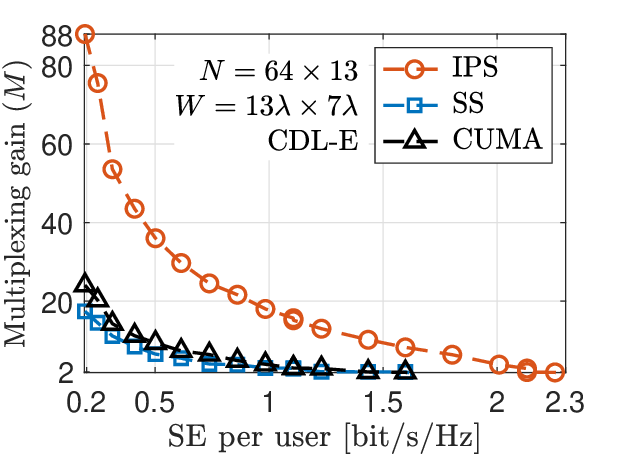}\label{subfig:MGvsSE_CDLChan_26GHz_NRF8}}
\caption{Multiplexing gain with different MCS under CDL channels. }\label{fig:MGvsSE_CDLChan}
\end{figure*}

\vspace{-2mm}
\subsection{Wideband Performance}
Now, we extend the system to a wideband system as in \cite{hong2025Downlink}, and perform simulations in CDL channels. The multiplexing gain results under CDL channels are presented in Fig.~\ref{fig:MGvsSE_CDLChan}. The results show consistent trends with those observed under finite-scattering channels. All schemes achieve higher multiplexing gain utilizing a robust MCS with lower SE per UT, and the multiplexing gain escalates with increase in the number of RF chains. Note that in wideband CDF channels, CUMA no longer exhibits superior performance at the frequency band of $26~{\rm GHz}$. This may be attributable to the channel variations among the subcarriers, which could induce inaccuracies in port selection and combination within the CUMA framework. Instead, slow FAMA with the proposed port selection methods achieves the optimal multiplexing gain in all examined cases. When $N_{\rm RF}=2$, the results of IPS and EPS are similar, indicating the effectiveness of IPS for balancing performance and complexity. The multiplexing gain of slow FAMA with IPS is particularly remarkable at the higher frequency, surpassing $20$ with a limited number of RF chains ($N_{\rm RF} = 2$), and reaching more than $80$ UTs when $N_{\rm RF} = 8$. Nonetheless, the multiplexing gain diminishes as SE per UT increases, ultimately falling below $2$ when ${\rm SE} > 2~{\rm bit/s/Hz}$.

\section{Conclusion}\label{sec:conclusion}
In this paper, we addressed the critical port selection problem in multi-port slow FAMA systems. We proposed three methods, namely EPS, IPS, and DPS, with their complexities analyzed and their performances evaluated. Simulation results demonstrated that the proposed IPS and DPS approach performance that is nearly identical to the optimal EPS. In scenarios with high-resolution FAS, IPS appears to be an effective option for balancing performance and complexity. Utilizing the proposed IPS method, the slow FAMA network can achieve approximately $10$ to $40$ multiplexing gains at the $6$ GHz frequency band and $20$ to $80$ multiplexing gains at the $26$ GHz frequency band, with a limited number of RF chains and a typical mobile phone-sized FAS at the receiver. This fundamental finding validated the feasibility of realizing significant connectivity benefits of multi-port slow FAMA, thereby providing an essential and practical pathway for its adoption in future wireless communication networks.

\vspace{-2mm}
\bibliographystyle{IEEEtran}

\begin{thebibliography}{10}
\providecommand{\url}[1]{#1}
\csname url@samestyle\endcsname
\providecommand{\newblock}{\relax}
\providecommand{\bibinfo}[2]{#2}
\providecommand{\BIBentrySTDinterwordspacing}{\spaceskip=0pt\relax}
\providecommand{\BIBentryALTinterwordstretchfactor}{4}
\providecommand{\BIBentryALTinterwordspacing}{\spaceskip=\fontdimen2\font plus
\BIBentryALTinterwordstretchfactor\fontdimen3\font minus
  \fontdimen4\font\relax}
\providecommand{\BIBforeignlanguage}[2]{{%
\expandafter\ifx\csname l@#1\endcsname\relax
\typeout{** WARNING: IEEEtran.bst: No hyphenation pattern has been}%
\typeout{** loaded for the language `#1'. Using the pattern for}%
\typeout{** the default language instead.}%
\else
\language=\csname l@#1\endcsname
\fi
#2}}
\providecommand{\BIBdecl}{\relax}
\BIBdecl
\bibitem{tariq-2020}
F. Tariq {\em et al.}, ``A speculative study on 6G,'' {\em IEEE Wireless Commun.}, vol. 27, no. 4, pp. 118--125, Aug. 2020.
\bibitem{andrews20246gtakes}
J. G. Andrews, T. E. Humphreys and T. Ji, ``6G takes shape,'' {\em IEEE BITS Inf. Theory Mag.}, vol. 4, no. 1, pp. 2--24, Mar. 2024.
\bibitem{ngo2024ultradense}
H. Q. Ngo, G. Interdonato, E. G. Larsson, G. Caire and J. G. Andrews, ``Ultradense cell-free massive MIMO for 6G: Technical overview and open questions,'' \emph{Proc. IEEE}, vol. 112, no. 7, pp. 805--831, Jul. 2024.
\bibitem{nguyen20226ginternet}
D. C. Nguyen {\em et al.}, ``6G internet of things: A comprehensive survey,'' {\em IEEE Internet Things J.}, vol. 9, no. 1, pp. 359--383, Jan. 2022.
\bibitem{silva2025distributed}
M. V. da Silva, E. Eldeeb, M. Shehab, H. Alves and R. D. Souza, ``Distributed learning methodologies for massive machine type communication,'' {\em IEEE Internet Things Mag.}, vol. 8, no. 1, pp. 102--108, Jan. 2025.
\bibitem{clercks2024multiple}
B. Clerckx {\em et al.}, ``Multiple access techniques for intelligent and multifunctional 6G: Tutorial, survey, and outlook,'' \emph{Proc. IEEE}, vol. 112, no. 7, pp. 832--879, Jul. 2024.
\bibitem{wang2024extremely}
Z. Wang {\em et al.}, ``Extremely large-scale MIMO: Fundamentals, challenges, solutions, and future directions,'' \emph{IEEE Wireless Commun.}, vol. 31, no. 3, pp. 117--124, Jun. 2024.
\bibitem{pereira2022anOverview}
F. A. Pereira de Figueiredo, ``An overview of massive MIMO for 5G and 6G,'' \emph{IEEE Latin America Trans.}, vol. 20, no. 6, pp. 931--940, Jun. 2022.
\bibitem{ahmed2024unveil}
A. Ahmed {\em et al.}, ``Unveiling the potential of NOMA: A journey to next generation multiple access,'' \emph{IEEE Commun. Surv. \& Tut.}, vol. 27, no. 5, pp. 3099--3164, Oct. 2025.
\bibitem{mao2022rsma}
Y. Mao {\em et al.}, ``Rate-splitting multiple access: Fundamentals, survey, and future research trends,'' \emph{IEEE Commun. Surv. \& Tut.}, vol. 24, no. 4, pp. 2073--2126, 4th Quart. 2022.

\bibitem{wong2022FAMA}
K. K. Wong and K. F. Tong, ``Fluid antenna multiple access,'' \emph{IEEE Trans. Wireless Commun.}, vol.~21, no.~7, pp. 4801--4815, Jul. 2022.
\bibitem{wong2020FAS}
K. K. Wong, K. F. Tong, Y. Zhang, and Z. Zheng, ``Fluid antenna system for {6G}: When {Bruce Lee} inspires wireless communications,'' {\em Elect. Lett.}, vol.~56, no.~24, pp.~1288--1290, Nov. 2020.
\bibitem{wong2021FAS}
K. K. Wong, A. Shojaeifard, K. F. Tong, and Y. Zhang, ``Fluid antenna systems,'' {\em IEEE Trans. Wireless Commun.}, vol. 20, no. 3, pp. 1950--1962, Mar. 2021.

\bibitem{New2024aTutorial}
W. K. New {\em et al.}, ``A tutorial on fluid antenna system for 6G networks: Encompassing communication theory, optimization methods and hardware designs,'' \emph{IEEE Commun. Surv. \& Tut.}, vol. 27, no. 4, pp. 2325--2377, Aug. 2025.
\bibitem{Lu-2025}
W.-J. Lu {\em et al.}, ``Fluid antennas: Reshaping intrinsic properties for flexible radiation characteristics in intelligent wireless networks,'' {\em IEEE Commun. Mag.}, vol. 63, no. 5, pp. 40--45, May 2025.
\bibitem{hong2025contemporary}
H. Hong {\em et al.}, ``A contemporary survey on fluid antenna systems: Fundamentals and networking perspectives,'' {\em  IEEE Trans. Netw. Sci. Eng.}, \url{doi:10.1109/TNSE.2025.3613225}, Sept. 2025.
\bibitem{New-2026jsac}
W. K. New {\em et al.}, ``Fluid antenna systems: Redefining reconfigurable wireless communications,'' {\em IEEE J. Select. Areas Commun.}, \url{doi: 10.1109/JSAC.2025.3632097}, 2026.

\bibitem{shen2024design}
Y. Shen {\em et al.}, ``Design and implementation of mmWave surface wave enabled fluid antennas and experimental results for fluid antenna multiple access,'' {\em arXiv preprint}, \url{arXiv:2405.09663}, May 2024.
\bibitem{Shamim-2025}
R. Wang {\em et al.}, ``Electromagnetically reconfigurable fluid antenna system for wireless communications: Design, modeling, algorithm, fabrication, and experiment,'' {\em IEEE J. Select. Areas Commun.}, \url{doi: 10.1109/JSAC.2025.3625163}, 2026.
\bibitem{Liu-2025arxiv}
B. Liu, K.-F. Tong, K. K. Wong, C.-B. Chae, and H. Wong, ``Programmable meta-fluid antenna for spatial multiplexing in fast fluctuating radio channels,'' {\em Optics Express}, vol. 33, no. 13, pp. 28898--28915, 2025.
\bibitem{Zhang-jsac2026}
S. Zhang {\em et al.}, ``Fluid antenna systems enabled by reconfigurable holographic surfaces: Beamforming design and experimental validation,'' {\em IEEE J. Select. Areas Commun.}, \url{doi: 10.1109/JSAC.2025.3618797}, 2026.
\bibitem{zhang2024pixel}
J. Zhang {\em et al.}, ``A novel pixel-based reconfigurable antenna applied in fluid antenna systems with high switching speed,'' {\em IEEE Open J. Antennas \& Propag.}, vol. 6, no. 1, pp. 212--228, Feb. 2025.
\bibitem{liu-2025iot}
B. Liu, T. Wu, K. K. Wong, H. Wong, and K. F. Tong, ``Wideband pixel-based fluid antenna system: An antenna design for smart city,'' to appear in {\em IEEE Internet of Things J.}, 2025.

\bibitem{Khammassi2023}
M. Khammassi, A. Kammoun and M.-S. Alouini, ``A new analytical approximation of the fluid antenna system channel,'' {\em IEEE Trans. Wireless Commun.}, vol. 22, no. 12, pp. 8843--8858, Dec. 2023.
\bibitem{espinosa2024anew}
P. Ram\'{i}rez-Espinosa, D. Morales-Jimenez and K. K. Wong, ``A new spatial block-correlation model for fluid antenna systems,'' \emph{IEEE Trans. Wireless Commun.}, vol. 23, no. 11, pp. 15829--15843, Nov. 2024.
\bibitem{New2023fluid}
W. K. New, K. K. Wong, H. Xu, K. F. Tong and C.-B. Chae, ``Fluid antenna system: New insights on outage probability and diversity gain,''  {\em IEEE Trans. Wireless Commun.}, vol. 23, no. 1, pp. 128--140, Jan. 2024.
\bibitem{new2023information}
W. K. New, K. K. Wong, H. Xu, K. F. Tong, and C.-B. Chae, ``An information-theoretic characterization of {MIMO-FAS}: Optimization, diversity-multiplexing tradeoff and \textit{q}-outage capacity,'' \emph{IEEE Trans. Wireless Commun.}, vol. 23, no. 6, pp. 5541--5556, Jun. 2024.
\bibitem{zhu2025fluid}
X. Zhu {\em et al.}, ``Fluid antenna systems: A geometric approach to error probability and fundamental limits,'' {\em arXiv preprint}, \url{arXiv:2509.08815}, Sept. 2025.
\bibitem{xu2023channel}
H. Xu {\em et al.}, ``Channel estimation for {FAS}-assisted multiuser {mmWave} systems,'' {\em IEEE Commun. Lett.}, vol.~28, no.~3, pp.~632--636, Mar. 2024.
\bibitem{new2025channel}
W. K. New {\em et al.}, ``Channel estimation and reconstruction in fluid antenna system: Oversampling is essential,'' {\em IEEE Trans. Wireless Commun.}, vol. 24, no. 1, pp. 309--322, Jan. 2025.
\bibitem{zhang2025successive}
Z. Zhang, J. Zhu, L. Dai and R. W. Heath, Jr., ``Successive Bayesian reconstructor for channel estimation in fluid antenna systems,'' {\em IEEE Trans. Wireless Commun.}, vol. 24, no. 3, pp. 1992--2006, Mar. 2025.

\bibitem{hong2025fasofdm}
H. Hong {\em et al.}, ``FAS meets OFDM: Enabling wideband 5G NR,'' {\em IEEE Trans. Commun.}, vol. 73, no. 11, pp. 12884--12898, Nov. 2025.


\bibitem{wong2023sFAMA}
K. K. Wong, D. Morales-Jimenez, K. F. Tong, and C. B. Chae, ``Slow fluid antenna multiple access,'' \emph{IEEE Trans. Commun.}, vol.~71, no.~5, pp. 2831--2846, May 2023.
\bibitem{Xu2024revisiting}
H.~Xu {\em et al.}, ``Revisiting outage probability analysis for two-user fluid antenna multiple access system,'' \emph{IEEE Trans. Wireless Commun.}, vol. 23, no. 8, pp. 9534--9548, Aug. 2024.
\bibitem{wong2022fast}
K. K. Wong, K. F. Tong, Y. Chen, and Y. Zhang, ``Fast fluid antenna multiple access enabling massive connectivity,'' \emph{IEEE Commun. Lett.}, vol.~27, no.~2, pp. 711--715, Feb. 2023.
\bibitem{Waqar2023deep}
N. Waqar, K. K. Wong, K. F. Tong, A. Sharples, and Y. Zhang, ``Deep learning enabled slow fluid antenna multiple access,'' \emph{IEEE Commun. Lett.}, vol. 27, no. 3, pp. 861--865, Mar. 2023.
\bibitem{hong20245gcoded}
H. Hong, K. K. Wong, K. F. Wong, H. Xu and H. Li, ``5G-coded fluid antenna multiple access over block fading channels,'' {\em Elect. Lett.}, vol. 61, no. 1, Jan. 2025.
\bibitem{hong2025coded}
H. Hong, K. K. Wong, K. F. Tong, H. Shin, and Y. Zhang, ``Coded fluid antenna multiple access over fast fading channels,'' \emph{IEEE Wireless Commun. Lett.}, vol.~14, no.~4, pp.~1249--1253, Apr. 2025.
\bibitem{waqar2025turbocharging}
N. Waqar, K. K. Wong, C.-B. Chae and R. Murch, ``Turbocharging fluid antenna multiple access,'' {\em IEEE Trans. Wireless Commun.}, \url{doi: 10.1109/TWC.2025.3607824}, 2025.
\bibitem{hong2025Downlink}
H. Hong {\em et al.}, ``Downlink OFDM-FAMA in 5G-NR systems,'' {\em IEEE Trans. Wireless Commun.}, \url{doi: 10.1109/TWC.2025.3577771}, Jun. 2025.


\bibitem{Wong2024cuma}
K. K. Wong, C. B. Chae, and K. F. Tong, ``Compact ultra massive antenna array: A simple open-loop massive connectivity scheme,'' {\em IEEE Trans. Wireless Commun.}, vol. 23, no. 6, pp. 6279--6294, Jun. 2024.
\bibitem{rao2025geometric}
C. Rao {\em et al.}, ``Geometric port selection in CUMA Systems,'' {\em arXiv preprint}, \url{arXiv:2509.20299}, Sept. 2025.
\bibitem{coma2024slow}
J. P. Gonz\'{a}lez-Coma and F. J. L\'{o}pez-Mart\'{i}nez, ``Slow fluid antenna multiple access with multiport receivers,'' {\em arXiv preprint}, \url{arXiv:2507.17505}, Jul. 2025.


\bibitem{buzzi2016clustered}
S.~Buzzi and C.~D'Andrea, ``On clustered statistical MIMO millimeter wave channel simulation,'' \emph{arXiv preprint}, \url{arXiv:1604.00648v2}, May 2016.
\bibitem{khandelwal2014new}
V. Khandelwal and Karmeshu, ``A new approximation for average symbol error probability over log-normal channels,'' {\em IEEE Wireless Commun. Lett.}, vol. 3, no. 1, pp. 58--61, Feb. 2014.
\bibitem{shah1998performance}
A. Shah and A. M. Haimovich, ``Performance analysis of optimum combining in wireless communications with Rayleigh fading and cochannel interference,'' {\em IEEE Trans. Commun.}, vol. 46, no. 4, pp. 473--479, Apr. 1998.
\bibitem{38214}
``{NR}; {P}hysical layer procedures for data,'' Available [Online]: \url{https://www.3gpp.org/ftp/Specs/archive/38_series/38.214/38214-i40.zip}, Last Accessed on 2024-09-23.
\bibitem{38901}
``Study on channel model for frequencies from 0.5 to 100 GHz,'' Available [Online]: \url{https://www.3gpp.org/ftp/Specs/archive/38_series/38.901/8901-i00.zip}, Last Accessed on 2024-04-03.


\end{thebibliography}

\end{document}